\documentclass[a4paper,11pt]{article}
\usepackage{jheppub} % for details on the use of the package, please
\usepackage{amsmath,amssymb,amsthm,amscd,graphicx}
\usepackage{psfrag}
\usepackage[english]{babel}
\usepackage{float}\input epsf.sty

\usepackage{color}
\newtheorem{theorem}{Theorem}[section]

\theoremstyle{definition}

\newtheorem{example}[theorem]{Example}

\newcommand{\CA}{{\cal A}}

\newcommand{\CC}{{\cal C}}
\newcommand{\CE}{{\cal E}}

\newcommand{\CG}{{\cal G}}

\newcommand{\CI}{{\cal I}}

\newcommand{\CO}{{\cal O}}

\newcommand{\CQ}{{\cal Q}}
\newcommand{\CR}{{\cal R}}

\newcommand{\CW}{{\cal W}}

\def\IR{{\mathbb R}}

\def\IP{{\mathbb P}}

\def\IF{{\mathbb F}}

\newcommand{\fad}{\operatorname{\Phi}_{\mathsf{b}}}

\newcommand{\tr}{{\rm Tr}}
\newcommand{\re}{{\rm e}}
\newcommand{\ri}{{\rm i}}
\newcommand{\rd}{{\rm d}}

\newcommand{\mO}{\mathsf{O}}

\newcommand{\mx}{\mathsf{x}}

\newcommand{\mm}{\mathsf{p}}

\newcommand{\mb}{{\mathsf{b}}}
\newcommand{\mH}{\mathsf{H}}

\newcommand{\be}{\begin{equation}}
\newcommand{\ee}{\end{equation}}
\newcommand{\ba}{\begin{aligned}}
\newcommand{\ea}{\end{aligned}}
\newcommand{\ben}{\begin{eqnarray}\displaystyle}
\newcommand{\een}{\end{eqnarray}}

\newcommand{\sectiono}[1]{\section{#1}\setcounter{equation}{0}}

\newdimen\tableauside\tableauside=1.0ex
\newdimen\tableaurule\tableaurule=0.4pt
\newdimen\tableaustep
\def\phantomhrule#1{\hbox{\vbox to0pt{\hrule height\tableaurule width#1\vss}}}
\def\phantomvrule#1{\vbox{\hbox to0pt{\vrule width\tableaurule height#1\hss}}}
\def\sqr{\vbox{%
  \phantomhrule\tableaustep
  \hbox{\phantomvrule\tableaustep\kern\tableaustep\phantomvrule\tableaustep}%
  \hbox{\vbox{\phantomhrule\tableauside}\kern-\tableaurule}}}
\def\squares#1{\hbox{\count0=#1\noindent\loop\sqr
  \advance\count0 by-1 \ifnum\count0>0\repeat}}
\def\tableau#1{\vcenter{\offinterlineskip
  \tableaustep=\tableauside\advance\tableaustep by-\tableaurule
  \kern\normallineskip\hbox
    {\kern\normallineskip\vbox
      {\gettableau#1 0 }%
     \kern\normallineskip\kern\tableaurule}%
  \kern\normallineskip\kern\tableaurule}}
\def\gettableau#1{\ifnum#1=0\let\next=\null\else
\squares{#1}\let\next=\gettableau\fi\next}

\newcommand{\figref}[1]{Fig.~\protect\ref{#1}}
\title{\huge{Quantum curves as quantum distributions}}

\author{Marcos Mari\~no and Szabolcs Zakany}

\affiliation{D\'epartement de Physique Th\'eorique et Section de Math\'ematiques\\
Universit\'e de Gen\`eve, Gen\`eve, CH-1211 Switzerland}

\emailAdd{Marcos.Marino@unige.ch, Szabolcs.Zakany@unige.ch} 

\abstract{Topological strings on toric Calabi--Yau threefolds can be defined non-perturbatively in terms of a non-interacting Fermi gas of $N$ particles. 
Using this approach, we propose a definition of quantum mirror curves as quantum distributions on phase space. The quantum distribution is obtained as the 
Wigner transform of the reduced density matrix of the Fermi gas. We show that the classical mirror geometry emerges in the strongly coupled, large $N$ limit in 
which $\hbar \sim N$. In this limit, the Fermi gas has effectively zero temperature, and the Wigner distribution becomes sharply supported on the interior of the 
classical mirror curve. The quantum fluctuations around the classical 
limit turn out to be captured by an improved version of the universal scaling form of Balazs and Zipfel. } 
\begin{document}
\maketitle

\sectiono{Introduction}

In quantum theories, classical geometric structures should be replaced by a suitable notion of quantum geometry. 
Although it is not clear how quantum geometry should be defined 
in general, we expect that classical geometry should be recovered in an appropriate classical limit. For small but finite $\hbar$, i.e. in a semiclassical approximation, the quantum geometry should display some sort of quantum fluctuations around the classical limit.

The problem of finding quantum correlates of classical geometric constructions appears already in elementary quantum mechanics. Let us consider 
for example a classical, conservative mechanical system in one dimension with a Hamiltonian $H(x,p)$. The curve in phase space defined by the submanifold of constant energy $E$, 
\be
\label{curve}
H(x, p)=E,  
\ee
can be regarded as the classical geometry of the system. What is the quantum version of this curve? The most obvious answer is to go into the world of operators and replace (\ref{curve}) by the time-independent Schr\"odinger equation 
\be
\mH \varphi= E\varphi, 
\ee
where $\mH$ is an appropriate quantization of the Hamiltonian. However, once we formulate the problem in terms of operators, it 
is not easy to recover the geometric intuitions of classical physics. In fact, one of the main problems in defining quantum geometry is precisely the 
conceptual mismatch between the world of operators on a Hilbert space, and the classical world of functions on phase space. 
One possibility to overcome this mismatch is to use Wigner's formulation of quantum theory (see \cite{hillery,zachos-treatise} for overviews). 
Wigner's formulation is based on quasi-probability distributions in phase space, and it is particularly useful to study semiclassical physics. In this formulation, the classical curve (\ref{curve}) should be understood as the limiting support of a quantum distribution, in a suitable classical limit. 

In theories of quantum gravity, in which the geometry itself should be quantized in one way or another, the problem of finding appropriate notions of 
quantum geometry is even more delicate. In this context, the language of quantum distributions has also proved to be a valuable tool. In 
\cite{halliwell,anderson,habib} the Wigner distribution was used to clarify the emergence of 
classical geometry from the Hartle--Hawking wavefunction of the universe \cite{hh}. In theories of quantum gravity with a 
dual quantum-mechanical description, quantum distributions in the dual theory have been also 
advocated as appropriate notions of quantum geometry in the gravitational theory. For example, in non-critical string theory, the Wigner distribution 
associated to the FZZT brane wavefunction has been studied in \cite{gomez,ambjorn} as a precise definition of the 
``quantum Riemann surface" underlying doubly-scaled matrix models 
\cite{moore-rs, mmss}\footnote{Quantum distributions have been also studied in the context of the $c=1$ string 
in \cite{dmw1,dmw2}.}. In type IIB superstring theory, certain classical backgrounds 
are encoded in two-dimensional curves which can be regarded as Fermi surfaces for a dual quantum-mechanical system of non-interacting fermions 
\cite{berenstein, llm}. The quantum distributions associated to these Fermi droplets have been 
used in \cite{babel} to understand the emergence of classical geometry from the quantum system (see \cite{st,joan1,joan2} for further 
developments along these lines). Since FZZT branes can be reinterpreted as 
fermions \cite{adkmv}, the quantum description of the geometry in these two string theory examples involves in a crucial way a non-interacting Fermi gas. 

In \cite{ghm} a non-perturbative description of topological strings on toric Calabi--Yau (CY) 
manifolds was proposed, based on a non-interacting Fermi gas. The one-body density matrix of the Fermi gas is obtained 
by quantizing the mirror curve to the toric CY. This description involves two parameters: $N$, the number of particles in the gas, 
and the Planck constant $\hbar$ (for simplicity, we restrict our discussion in this paper to mirror curves of genus one). It was 
conjectured in \cite{ghm,mz} that the canonical partition function $Z(N, \hbar)$ of this Fermi gas provides a non-perturbative definition 
of the topological string free energy. More precisely, in the 't Hooft-like limit
\be
\label{sc-limit}
N, \hbar  \rightarrow \infty, \qquad {N \over \hbar}=\lambda \, \, \, {\rm fixed}, 
\ee
one has 
\be
\label{as-zn}
\log Z(N, \hbar) \sim \sum_{g\ge 0} \hbar^{2-2g} \mathcal F_g(\lambda), 
\ee
where $\mathcal F_g(\lambda)$ is the genus $g$ free energy of the topological string, and $\lambda$ is identified with a flat coordinate of the CY moduli space. 
Note that, according to this asymptotic expansion, the topological string coupling constant is proportional to $1/\hbar$, and the limit (\ref{sc-limit}) is in fact the classical limit of the topological string. 

It is natural to use this Fermi gas formulation as a tool to explore ``stringy" notions of geometry. In topological string theory 
on toric CY manifolds, the classical geometry is the classical mirror curve, which is the target geometry of the B-model and 
can be regarded as the analogue of the curve (\ref{curve}). There has been a 
lot of activity in the last years in the search for ``quantum" versions of the algebraic curves appearing in topological string theory and in related 
contexts (see for example \cite{dm1, dm2} for overviews, and \cite{be-wkb,ms} for some recent developments). In this paper, 
we will provide a non-perturbative definition of quantum mirror curves based on a quantum distribution on phase space associated to the Fermi gas. As in \cite{babel}, we will 
use the reduced one-particle density matrix of the quantum gas, and the associated Wigner distribution, as 
precise definitions of the quantum geometry. Such 
a definition must lead to the appropriate classical curve in the ``classical" limit (\ref{sc-limit}), and should display quantum fluctuations around it. This is far from obvious, since (\ref{sc-limit}) is not the standard  classical limit of the Fermi gas, in which $\hbar\rightarrow 0$. However, we will give evidence that, in the limit (\ref{sc-limit}), the Wigner distribution associated to a quantum mirror curve has 
the sharp, step function shape typical of Fermi gases at zero temperature. More precisely, it 
becomes constant inside the classical mirror curve, and vanishes outside. The complex structure of the mirror curve 
turns out to be determined by the 't Hooft parameter $\lambda$ through the mirror map. 
In this way, the classical mirror curve emerges, in the semiclassical limit, as the boundary of the support of the Wigner function. The underlying reason for 
this behavior is the remarkable 
duality structure of quantum mirror curves \cite{ghm,wzh,hatsuda-comments}, which is inherited from the modular duality of Weyl operators \cite{faddeev}. Thanks to this duality, the 
limit (\ref{sc-limit}) behaves both as a classical limit, and as a zero temperature limit. 

The sharp, strict classical limit of quantum distributions is smoothed out at small but finite $\hbar$. In this regime, Wigner distributions are described by 
universal functions which provide a precise description of the leading quantum fluctuations around the classical limit (at least in one dimension). In the case of the 
Wigner distribution of highly excited eigenstates, the corresponding 
function is an Airy function, as first found by Berry in \cite{berry}. For the Wigner distribution associated to the reduced density matrix of a Fermi gas at 
zero temperature, the universal function is an integral of the Airy function, as first found by Balazs and Zipfel in \cite{bzipfel}. In this paper we will propose a slightly improved 
version of the Balazs--Zipfel approximation, and we will argue that it describes the semiclassical regime of the Wigner distributions associated to quantum mirror curves.

This paper is organized as follows. In section \ref{sec-gen-fermi} we review general aspects of non-interacting Fermi gases, focusing on the reduced density matrix, 
the corresponding Wigner distribution, and their classical limit. In section \ref{sec-qmc} we apply these tools to the Fermi gases associated to 
quantum mirror curves. We argue that the Wigner distribution associated to the reduced density matrix provides a 
precise, quantitative definition of the quantum geometry associated to a mirror curve. In particular, we make a general conjecture about the classical limit 
of this Wigner distribution, in which the classical mirror curves emerges as the boundary of its support, and we provide evidence for it. We conclude with some open problems 
and suggestions for future research. The Appendices are devoted to detailed derivations of semiclassical limits of Wigner distributions associated to Fermi gases. In 
Appendix \ref{ho} we consider fermions in a harmonic potential, and we obtain an improved version of the Balazs--Zipfel approximation in this case. In Appendix 
\ref{app-localF0} we perform a similar but more involved calculation in the case of local $\IF_0$, providing in this way a direct verification of our conjecture for this geometry.

\sectiono{Fermions, quantum distributions, and the classical limit}
\label{sec-gen-fermi}

\subsection{The thermodynamics of non-interacting Fermi gases}

 Let us consider a system of $N$ one-dimensional fermions described by the density matrix $\rho_N$. 
 Our basic observable will be the {\it reduced density matrix}, or {\it one-particle correlation function}, defined by 
\be
\label{cn-gen}
C_N(x,y)={1\over N} \tr \left( \rho_N \, \hat \psi^\dagger(y) \hat \psi(x)\right), 
\ee
where $\hat \psi^\dagger (x)$, $\hat \psi(x)$ are standard many-body creation/annihilation operators for fermions in the position state $|x\rangle$ (see 
e.g. \cite{negele-orland}). Equivalently, one has 
\be
C_N(x, y)= \int_{\IR^{N-1}} \rho_N \left( x, x_1, \cdots, x_{N-1}; y, x_1, \cdots, x_{N-1} \right)\, \rd x_1 \cdots \rd x_{N-1}. 
\ee
We note that the reduced density matrix satisfies the normalization condition, 
\be
\int_\IR C_N(x,x) \, \rd x =1. 
\ee

Let us now consider a system of $N$ identical, non-interacting fermions, with total Hamiltonian 
\be
\label{total-H}
\mH_N= \sum_{i=1}^N \mH (i), 
\ee
where $\mH (i)$ is the one-body Hamiltonian $\mH$ for the $i$-th particle. We will assume that $\mH$ has a discrete, infinite spectrum, as it happens 
for example for a confining potential. In the case of non-interacting fermions at zero temperature, the one-particle correlation function can be easily computed. 
Let $\{ |\varphi_n\rangle\}_{n=0,1,2, \cdots}$ be 
an orthonormal basis for 
the one-particle Hilbert space, made out of eigenfunctions of the one-body Hamiltonian, i.e. 
\be
\mH |\varphi_n\rangle= E_n  |\varphi_n\rangle. 
\ee
The ground state of the $N$-fermion system is a Slater determinant, 
\be
\label{slater}
|\Psi_0 \rangle={1\over {\sqrt{N!}}} \sum_{\sigma \in S_N} (-1)^{\epsilon(\sigma)} |\varphi_{\sigma(1)}\rangle \otimes \cdots \otimes 
|\varphi_{\sigma(N)}\rangle,
\ee
where $S_N$ is the permutation group of $N$ elements and $\epsilon(\sigma)$ is the parity of $\sigma \in S_N$. 
The density matrix describing the ground state at $T=0$ is 
\be
\rho_N=| \Psi_0 \rangle \langle \Psi_0 |. 
\ee
The reduced density matrix is in this case, 
\be
C_N(x,y)={1\over N} \langle \Psi_0 | \hat \psi^\dagger(y) \hat \psi(x)|\Psi_0 \rangle. 
\ee
A simple calculation gives 
\be
\label{cnxy} 
C_N (x,y)={1\over N}  \sum_{n=0}^{N-1} \varphi^*_n(y)\varphi_n(x),  
\ee
where $\varphi_n(x)= \langle x | \varphi_n\rangle$. 

\begin{example} In the case of $N$ non-interacting fermions in a harmonic potential, (\ref{cnxy}) can be computed in a compact form (see for example \cite{dlms}). The classical Hamiltonian is, in appropriate units, 
\be
\label{cho}
H(x,p)={p^2 \over 2}+ {x^2 \over 2}, 
\ee
and the eigenfunctions of the quantum Hamiltonian are given by 
\be
\label{eigen-ho}
\psi_n(x)={1\over {\sqrt{2 \pi h_n}}} \re^{-{x^2\over 2 \hbar }} p_n(x), \qquad n=0,1,2,\cdots,
\ee
where 
\be
\label{pGauss}
p_n(x)=\left({\hbar \over 4}\right)^{n/2} H_n (x/{\sqrt { \hbar}}), \qquad h_n= {1\over {\sqrt{2 \pi}}} n! \left( {\hbar \over 2}\right)^{n+1/2}, 
\ee
and $H_n(x)$ is a Hermite polynomial. One can then use the Christoffel--Darboux formula to obtain 
\be
\label{cn-ho}
C_N(x, y)={1\over N}  \sum_{n=0}^{N-1} \psi_n(x)\psi_n(y) = {\sqrt{ \hbar \over 2 N }} {\psi_{N}(x) \psi_{N-1}(y) - \psi_{N}(y) \psi_{N-1}(x) \over x-y}.
\ee
\qed
\end{example}

Let us now consider a system of $N$ non-interacting fermions at finite temperature $T$. We will work in the canonical formalism. 
The {\it unnormalized reduced density matrix} at inverse temperature $\beta=(k_B T)^{-1}$ is defined as 
\be
 \rho^{(N)}_1 (x,y ;\beta) = \tr\left( \re^{-\beta \mH_N} \hat \psi^{\dagger}(y )  \hat \psi(x)\right).
 \ee
 It is related to the one-particle density matrix $C_N(x,y)$ in (\ref{cn-gen}) by a normalization factor, 
\be
C_N(x,y)= {1\over N Z_N} \rho^{(N)}_1 (x,y ;\beta).
\ee
We will be interested in the non-interacting case, in which the total Hamiltonian $\mH_N$ is of the form (\ref{total-H}). 
In this case, the canonical partition function is given by  
\be
\label{zn-int}
	Z_N =  \frac{1}{N!} \int_{\mathbb R^N} \rd x_1 \cdots \rd x_N  \,\, {\det}_N \rho(x_i,x_j; \beta).
\ee
In this equation, we have denoted by 
\be
\rho(x, y; \beta)=\langle x| \re^{-\beta \mH} |y \rangle=\sum_{k=0}^\infty \varphi_k(x) \re^{-\beta E_k} \varphi^*_k (y)
\ee
the integral kernel of the unnormalized, one-body density matrix $\re^{-\beta \mH}$. 
In the non-interacting case, the reduced density matrix can be obtained from Landsberg's recursion relation (see e.g. \cite{krauth}), 
\be
\rho^{(N)}_1(x,x';\beta)= \sum_{\ell=1}^N (-1)^{\ell-1} \rho(x,x'; \ell \beta) Z_{N-\ell}. 
\ee
Using the relations above, we obtain a useful representation for the reduced density matrix, 
\be
\label{cn-ft}
C_N(x,y)= \sum_{k=0}^\infty c_k^{(N)} \varphi_k(x) \varphi^*_k (y), 
\ee
where the coefficients $c_k^{(N)}$ are given by 
\be 
c_k^{(N)}={1\over N Z_N} \sum_{\ell=1}^N (-1)^{\ell-1} \re^{-\ell \beta E_k} Z_{N-\ell}. 
\ee
Note that
\be
\sum_{k=0}^N c_k^{(N)}=1. 
\ee
We can also write these coefficients in terms of the occupation numbers $n_k$ of the energy levels (see e.g. \cite{dlms}), 
\be
\label{ckN}
c_k^{(N)}= {1\over N Z_N} \sum_{ \{  n_\ell\} } n_k \re^{-\beta \sum_\ell n_\ell E_\ell}. 
\ee
From this expression, it is obvious that, in the limit of zero temperature, the $c_k^{(N)}$ have the following behavior 
\be
\label{ckn-lowT}
c_k^{(N)}\rightarrow \begin{cases} 1/N &\text{if $0\le k \le N-1$}, \\
0 &\text{if $ k\ge N$}. 
\end{cases}
\ee
Therefore, (\ref{cn-ft}) is the generalization of (\ref{cnxy}) to the finite temperature case.

It is also useful to work in the grand canonical formalism, so we introduce the grand-canonical reduced density matrix 
\be
\rho_1^{\rm GC}(x,x';\kappa)=\sum_{N=1}^{\infty} \rho^{(N)}_1(x,x';\beta) \kappa^N. 
\ee
From Landsberg's recursion, one finds
\be
\rho_1^{\rm GC}(x,x';\kappa)=\Xi (\kappa) \,  \left \langle x\left| {1\over \kappa^{-1} \re^{\beta \mH}+1} \right|x'\right\rangle, 
\ee
where
\be
\Xi(\kappa)= 1+\sum_{N\ge 1} Z_N \kappa^N
\ee
is the grand canonical partition function. 
We can now use some standard formulae in Fredholm theory to give alternative expressions for these quantities. If we regard $\mO=\re^{\beta \mH}$ as an operator, 
its resolvent is essentially the grand-canonical reduced density matrix: 
\be
R(x,x';\kappa, \beta)=\left \langle x\left| {1\over \re^{\beta \mH}+\kappa} \right|x'\right\rangle= {1\over \kappa \Xi(\kappa)}  \rho_1^{\rm GC}(x,x';\kappa, \beta). 
\ee
On the other hand, the resolvent in Fredholm theory is given by (see e.g. \cite{mz-wv})
\be
{1\over \Xi(\kappa)} \sum_{N \ge 0} \kappa^N B_N(x,x'; \beta), 
\ee
where
\be
\label{bn-integral}
B_N(x,x'; \beta)= {1\over N!} \int_{\IR^N} \rho \begin{pmatrix} x & x_1 & \cdots & x_N \\
x'& x_1 & \cdots &x_N \end{pmatrix}\rd x_1 \cdots \rd x_N. 
\ee
In this expression, the integrand is given by a determinant, 
\be
\rho \begin{pmatrix} x & x_1 & \cdots & x_N \\
x'& x_1 & \cdots &x_N \end{pmatrix}={\rm det}\left[ \rho(z_i, w_j;\beta) \right]_{i,j=0, 1, \cdots, N}, 
\ee
with
\be
\label{coords}
\ba
z_0=x, \quad z_i=x_i, \quad i=1,\cdots, N, \\
w_0=x', \quad w_i=x_i, \quad i=1,\cdots, N. 
\ea
\ee
Note that 
\be
B_0(x, x';\beta)= \rho(x, x';\beta). 
\ee
Therefore, 
\be
 \rho^{(N)}_1(x,x';\beta)= B_{N-1}(x,x';\beta), 
 \ee
and as a consequence, 
\be
\label{cn-bn}
C_N(x,y)={1\over N Z_N }  B_{N-1}(x,y;\beta).
\ee

\subsection{Classical and quantum geometry in Fermi gases}

Let us now come back to the problem mentioned in the introduction: how does one define a natural notion of ``quantum geometry", in such a 
way that classical geometry is recovered in some appropriate limit? We will focus on one-dimensional problems, in which the classical Hamiltonian is given by a function 
$H(x,p)$, and correspondingly the classical geometry is defined by a curve in a two-dimensional phase space as in (\ref{curve}). How can one define 
a ``quantum geometry" associated to such a curve? One way to do so is to construct a quasi-probability distribution in 
phase space which becomes ``localized" on the classical curve $H(x,p)=E$ in the classical limit. We 
recall that, given an eigenstate $\varphi_n(x)$ of the quantum Hamiltonian 
$\mH$ obtained from (\ref{curve}), the corresponding Wigner distribution is defined by 
\be
\label{wf}
f_{n}(x,p)= {1\over 2 \pi \hbar} \int_{\IR}  \varphi_n^* \left( x +{y \over 2}\right) \varphi_n\left(x- {y \over 2} \right) \re^{{\ri \over \hbar} p y} \rd y.
\ee
To understand the semiclassical limit of the Wigner distribution, we have to look at highly excited states, as expected from WKB analysis. In the leading WKB 
approximation, the energy levels are given by the Bohr--Sommerfeld approximation
\be
\label{bs-condition}
I(E)=  \hbar \left(n+{1\over 2}\right), \qquad n=0, 1,\cdots 
\ee
In this equation, $I(E)$ is the classical action variable, which is obtained as follows. Let $\CR$ be the region in phase space inside the curve (\ref{curve}), and let ${\cal C}$ be a path along the boundary of $\CR$. 
Then, 
\be
\label{iaction}
I(E) = {1\over 2 \pi} \oint_{\cal C} p(x) \rd x, 
\ee
and it is proportional to the volume of the region $\CR$. Let us now consider the double-scaling limit
\be 
\label{semi-lim}
n  \rightarrow \infty, \qquad \hbar \rightarrow 0, \qquad n  \hbar = \xi\qquad \text{fixed}. 
\ee
In this limit, the Bohr--Sommerfeld approximation becomes exact and defines implicitly a function $E(\xi)$ through
\be
\label{bs}
I(E)=\xi.  
\ee
It turns out that the Wigner function becomes in this limit a delta function distribution concentrated on the classical curve (\ref{curve}) \cite{voros-asymptotic,berry}
\be
\label{delta-limit}
f_n(x, p) \rightarrow {1\over 2 \pi} \delta(H(x,p)-E(\xi)). 
\ee
This limit only holds in the sense of distributions, against integration of appropriate functions (see \cite{ripamonti} for a detailed analysis 
of this limit in the case of the harmonic oscillator). In fact, the Wigner function approaches a delta function in a highly non-trivial way: it decays 
very fast outside the classical curve, it has a peak approximately 
at the classical curve, and it oscillates very rapidly around zero inside the curve. At small but nonzero $\hbar$, 
this non-trivial structure is captured by a universal limiting form which was derived by Berry in \cite{berry}. 
To state the result of \cite{berry}, let us assume that the region $\CR$ in phase space enclosed by the curve (\ref{curve}) 
is simply connected and convex. The equation for the curve defines locally a function $p(x)$. Given an arbitrary point $(x,p)$ in phase space, 
we obtain a value of $E$ through (\ref{curve}), therefore 
a value of $I$ through (\ref{iaction}). This defines the function $I(x,p)=I(H(x,p))$.

\begin{figure}-
\begin{center}
	\includegraphics[width=0.5\textwidth]{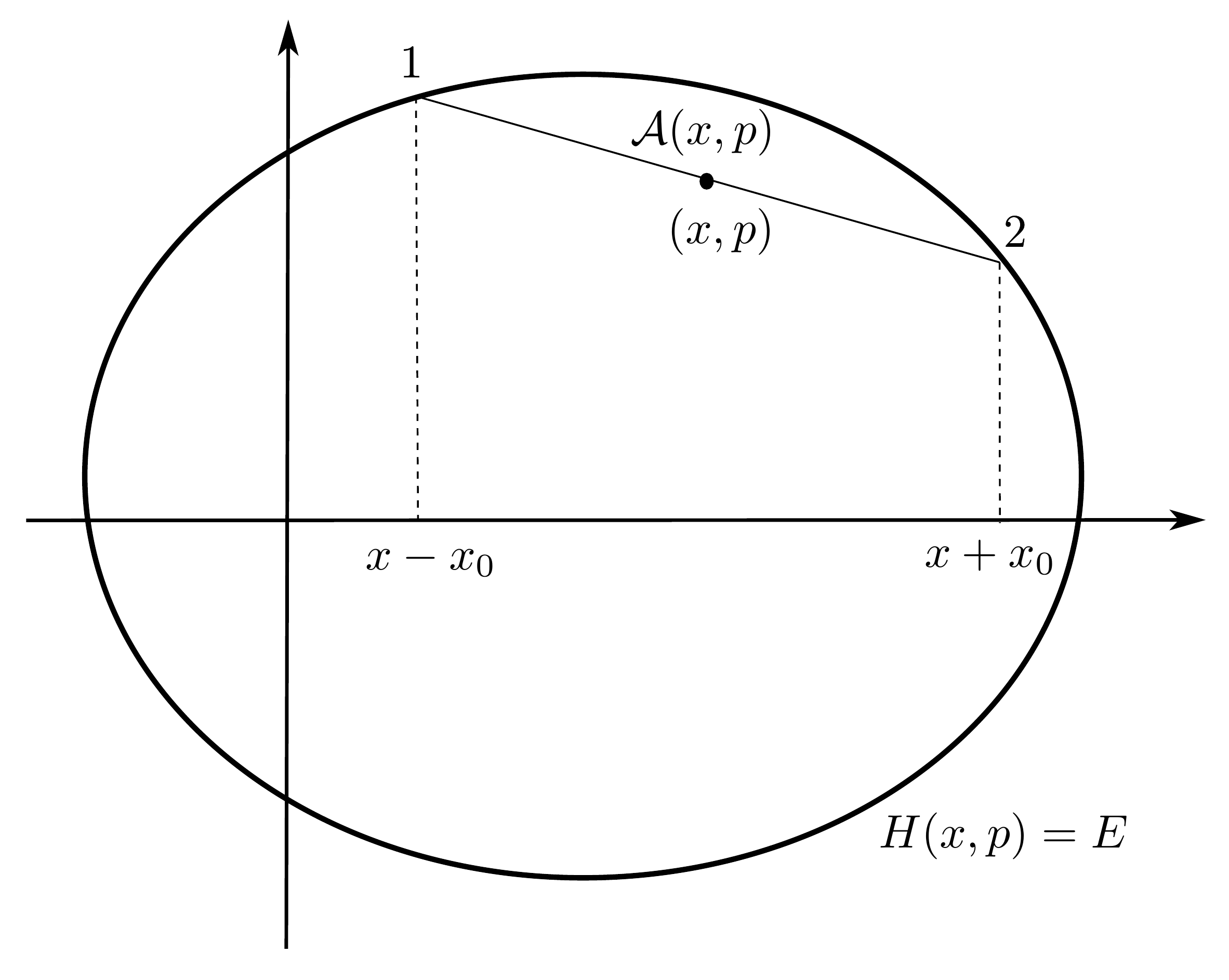} 
	\caption{Berry's chord construction. The point $(x,p)$ is the midpoint of a segment (or chord) whose endpoints lie on the 
	classical curve $H(x,p)=E$. The area of the region between the chord and the curve is denoted by $\CA(x,p)$ and we refer to it as the area of the chord. \label{fig3}}
\end{center}
\end{figure}

Suppose now that we fix $E$. For an arbitrary point $(x,p)$ inside the region $\mathcal R$, we define $x_0$ to be a solution of
\be
	\label{x0def}
	p(x+x_0)+p(x-x_0)=2p.
\ee
If $x_0$ is a solution, $-x_0$ is a solution, too. We obtain in this way two points
\be
\label{twopoints}
\ba
	(1) = (x-x_0,p(x-x_0)), \\
	(2) = (x+x_0,p(x+x_0)),
\ea
\ee
which lie on the curve, see \figref{fig3}. They are both at equal distance of the point $(x,p)$. The segment joining these two points is usually called the 
chord through the point $(x,p)$. The area of the region between the chord and the curve will be called the {\it chord area}, and we will denote it by ${\cal A}(x,p)$. 
We also define
\be
	\Delta_{1,2}(x,p) = \partial_{x}I(1) \partial_{p}I(2)-\partial_{x}I(2) \partial_{p}I(1), 
\ee
where $(1)$ and $(2)$ label the two points (\ref{twopoints}). Berry's formula for the uniform, semiclassical approximation to the Wigner function is given by 
\be
\label{berry-f}
	f_n (x,p) \approx \frac{\sqrt{2}}{\pi \hbar^{2/3}} \frac{(\frac{3}{2}\CA(x,p))^{1/6}}{\sqrt{\Delta_{1,2}(x,p)}} \, {\rm Ai} \left [ -\left ( \frac{3 \CA(x,p)}{2\hbar} \right )^{2/3} \right ], 
\ee
where ${\rm Ai}(z)$ denotes the Airy function. 
The quantum number $n$ enters the r.h.s. through the energy $E_n$, which should satisfy the Bohr--Sommerfeld quantization condition (\ref{bs-condition}). 
In principle, (\ref{berry-f}) is valid for points $(x,p)$ inside $\CR$, where the geometric chord construction makes sense. However, one can analytically 
 continue the function $\CA(x,p)$ to points outside $\CR$, where it becomes a complex number with phase $3 \pi/2$, in such a way that the argument of the 
 Airy function is positive. 
  
 When $(x,p)$ is near the classical curve (\ref{curve}), one can further approximate (\ref{berry-f}) by the so-called ``transitional 
 approximation," given by
\be
\label{wig-inter}
f_n (x,p) \approx {1\over \pi} \left( {1\over \hbar^2 B(x,p)}\right)^{1/3} {\rm Ai}\left[ 2 {I(x,p)-I(E_n)) \over \hbar^{2/3} B^{1/3}(x,p)} \right]. 
\ee
Here, 
\be\label{bxp}
B(x,p)= I_p^2 I_{xx}+I_x^2I_{pp} -2 I_p I_x I_{px}. 
\ee
The formula (\ref{wig-inter}) gives a universal scaling form for the Wigner function near the classical curve. From (\ref{wig-inter}), (\ref{delta-limit}) follows. 

The double-scaling limit (\ref{semi-lim}) is mathematically well-defined, but it would be nice to implement it physically. 
One way to achieve this is to consider a 
system of $N$ non-interacting fermions at zero temperature, 
with one-particle Hamiltonian $\mH(\mx, \mm)$. The Fermi exclusion principle guarantees that, in the thermodynamic limit in which 
$N$ is large, the edge of the Fermi sea will be in a highly excited state. The appropriate quantum distribution describing the Fermi gas 
is the Wigner transform of the reduced one-particle density 
matrix (\ref{cn-gen}):
\be
\label{wigner-reduced}
\CW_N(x,p)= {1\over 2 \pi \hbar} \int_\IR  C_N\left( x- {y \over 2}, x + {y \over 2} \right) \re^{\ri p y/\hbar} \rd y. 
\ee
At zero temperature, this distribution can be evaluated directly from (\ref{cnxy}) as a sum of Wigner functions, 
\be
\label{wr-zero}
\CW_N(x,p)= {1\over N}\sum_{n=0}^{N-1} f_n (x,p). 
\ee
Let us now consider the following double-scaling limit
\be
\label{ds-largeN}
N \rightarrow \infty, \quad \hbar \rightarrow 0, \qquad N \hbar = \xi_F \qquad \text{fixed}. 
\ee
This combines the thermodynamic limit $N \rightarrow \infty$ with the semiclassical limit of Quantum Mechanics $\hbar \rightarrow 0$. 
Using again the Bohr--Sommerfeld quantization condition, the limit (\ref{ds-largeN}) defines a Fermi energy $E_F$ as a function of $\xi_F$, 
\be
I(E_F)=  \xi_F. 
\ee
In this limit, the distribution at zero temperature (\ref{wr-zero}) becomes a constant in the region inside the classical curve $H(x, p)=E_F$, and zero outside, i.e. 
\be
\label{cl-fermi}
\CW_N(x,p) \approx {1\over 2 \pi I(E_F)} \Theta (E_F-H(x,p)), 
\ee
where $\Theta$ is the Heaviside step function. This result goes back to the Thomas--Fermi approximation for fermionic systems. A recent derivation 
can be found in \cite{dlms2} (we note however that in \cite{dlms2} the Fermi energy is fixed by normalization of the distribution, while in our case we use the Bohr--Sommerfeld quantization condition).

The limiting behavior (\ref{cl-fermi}) shows that, in a non-interacting Fermi gas at zero temperature, there is a 
natural definition of ``quantum geometry" based on the Wigner distribution associated to the reduced one-particle density matrix. 
The classical curve in phase space emerges in the double-scaling limit (\ref{ds-largeN}) as the boundary of the support of the distribution. 
Note that, at finite temperature, the Heaviside behavior in (\ref{cl-fermi}) is smoothed out by thermal fluctuations \cite{dlms2}. 
The strict classical limit of the geometry is only achieved at zero temperature. 

As in the case of the Wigner function associated to a highly excited state, the limit (\ref{cl-fermi}) occurs in a non-trivial way: outside the classical curve, the distribution decays rapidly, while inside the curve we have oscillations around its average value $(2 \pi I(E_F))^{-1}$. 
It is natural to look for analogues of (\ref{berry-f}) and (\ref{wig-inter}) for the quantum distribution 
(\ref{wr-zero}). The first result along this direction was obtained by Balazs and Zipfel in 1973 in \cite{bzipfel}. The Balazs--Zipfel scaling 
form is valid near the classical curve, and it is given by an integrated Airy function, 
\be
\label{cnapprox}
\CW_N(x,p) \approx{1 \over 2 \pi I(E_F) } \CI(t_{\rm BZ} ), 
\ee
where
\be
\label{airy-int}
\ba
\CI(z)&=\int_{z}^\infty {\rm Ai}(t) \rd t \\
		&= \frac{ {3^{1/3}} z }{\Gamma
   \left(-\frac{1}{3}\right)}\,
   _1F_2\left(\frac{1}{3};\frac{2}{3},\frac{4}{3};\frac{z^3}
   {9}\right)+\frac{ {3^{1/6}} z^2 \Gamma
   \left(\frac{2}{3}\right) }{4 \pi }\,
   _1F_2\left(\frac{2}{3};\frac{4}{3},\frac{5}{3};\frac{z^3}
   {9}\right)+\frac{1}{3},  
\ea
\ee
and the argument $t_{\rm BZ}$ is 
\be
\label{tstar}
t_{\rm BZ}= \left( {2 \over  \hbar}\right)^{2/3} {I(x,p)- I(E_F) \over I(E_F)^{1/3}}.
\ee
However, we have found in some examples that this result can be upgraded to a uniform 
approximation involving Berry's chord construction. The {\it improved Balazs--Zipfel 
approximation} to the Wigner distribution is given by,  
\be
\label{ibz}
\CW_N(x,p) \approx{1 \over 2 \pi I(E_F) } \mathcal I \left (-\left ( \frac{3\mathcal A(x,p)}{2\hbar} \right )^{2/3} \right ), 
\ee
where $\CA(x,p)$ is the area of the chord associated to the classical curve $H(x,p)= E_F$. 
In the improved version, the scaling function remains the same, but the argument $t_{\rm BZ}$ changes. 
In Appendix \ref{ho} we derive this improved result in the case of the harmonic oscillator. It is easy to check that, 
as $\hbar \rightarrow 0$, (\ref{ibz}) gives back (\ref{cl-fermi}). One can derive from (\ref{ibz}) a ``transitional approximation" near 
the classical curve, as in 
(\ref{wig-inter}), which reads, 
\be
\label{ibz-trans}
 \CW_N(x,p) \approx{1 \over 2 \pi I(E_F) } \mathcal I \left( 2 {I(x,p)-I(E_F)) \over \hbar^{2/3} B^{1/3}(x,p)} \right). 
 \ee
 In the case of the harmonic oscillator, this transitional approximation agrees with the original result of Balasz and Zipfel, but in general they are different. In the examples we have considered, 
 (\ref{ibz}) gives a better match than (\ref{ibz-trans}), which in turn is better than (\ref{cnapprox}).

\sectiono{Quantum mirror curves as quantum distributions}
\label{sec-qmc}
\subsection{Topological strings and their classical limit} 
\label{sec-ts-sl}

A Fermi gas approach to topological strings on toric CY threefolds was proposed in \cite{ghm}, building on previous works \cite{adkmv,ns, km, wzh}. 
We will summarize here some basic ingredients of the theory, referring to the original paper 
\cite{ghm} and the review \cite{mmrev} for further details. For simplicity we will focus on 
toric CY threefolds whose mirror curve has genus one. In this case, the mirror curve is encoded in the equation 
\be
\label{mirror-curve}
\CO (\re^x, \re^p) =\kappa, 
\ee
where $\CO(\re^x, \re^p)$ is a polynomial in the exponentiated variables. This curve can be quantized by promoting $x$ and $p$ to canonically conjugate Heisenberg operators on $L^2(\IR)$, 
\be
[\mx, \mm]= \ri \hbar. 
\ee
Ordering ambiguities are resolved by using Weyl quantization. The polynomial $\CO(\re^x, \re^p)$ becomes an operator $\mO$, and its inverse
\be
\label{rho-def}
\rho= \mO^{-1}
\ee
turns out to be a trace class, self-adjoint operator on $L^2(\IR)$ (this requires positivity conditions on the parameters appearing in $\mO$, although the 
theory can be extended to more general values of the parameters \cite{cgum}). In particular, it is natural to regard $\rho$ as a canonical density matrix for a quantum Hamiltonian, i.e. 
\be
\label{rhoH}
\rho= \re^{- \mH}. 
\ee
The inverse temperature $\beta$ is set to one, although, as we will see in a moment, there is a natural notion of low temperature limit. 

The spectral problem for quantum mirror curves has been studied in detail in the last few years. In \cite{ghm}, a conjectural, exact expression for the spectral 
determinant of $\rho$ (or, equivalently, for the grand canonical partition function of the corresponding Fermi gas) 
was proposed. Exact quantization conditions for the spectrum can then be obtained from the vanishing locus of the spectral determinant. A useful formulation of these quantization 
conditions was proposed in \cite{wzh}, and it was later shown in \cite{ggu} that the formulations of \cite{ghm} and \cite{wzh} 
are equivalent (see \cite{swh,h-blowup} for further work along this direction).  
According to \cite{wzh}, the exact quantization condition for genus one geometries can be written as
 \be
 \label{eqc}
 {r C t^2 \over2} + B(\hbar)+ \hbar \left( f_{\rm NS}\left( t, \hbar \right) + f_{\rm NS} \left( {2 \pi t \over \hbar}, \frac{4\pi^2}{\hbar} \right) \right)= 2 \pi \hbar \left( n+{1\over 2} \right), \qquad n=0,1,2, \cdots
 \ee
In this equation, 
 \be
 B(\hbar) = B \left( 1+ {\hbar^2 \over 4 \pi^2} \right), 
 \ee
$C$, $r$, and $B$ are constant coefficients depending on the geometry under consideration, 
and $t$ is related to the energy through the so-called quantum mirror map \cite{acdkv}: 
 \be
 t= t(E, \hbar). 
 \ee
 The equation (\ref{eqc}) determines the energy levels $E_n$, $n=0,1,\cdots$ of the Hamiltonian $\mH$ defined by (\ref{rhoH}). 
 We note from (\ref{mirror-curve}) that the energy $E$ is related to the modulus $\kappa$ appearing in the equation of the mirror curve as
 \be
 \label{kappa-E}
 \kappa=\re^E. 
 \ee
 When $\hbar \rightarrow 0$, the quantum mirror map becomes the classical mirror map $t=t(E)$ relating the K\"ahler parameter to the modulus $\kappa$. 
 We also note that, for large $E$ and fixed $\hbar$, the quantum mirror map behaves as
 \be
 t (E, \hbar) = r E +\CO(\re^{-rE}). 
 \ee
  Finally, the function $f_{\rm NS}(t, \hbar)$ can be expressed in terms of the Nekrasov--Shatashvili (NS) limit $F_{\rm NS}(t, \hbar)$
 of the refined topological string free energy, as, 
\be
f_{\rm NS}(t, \hbar)= r {\partial F_{\rm NS}^{\rm inst} \over \partial t}, 
\ee
where the superscript indicates that we only keep the instanton part of the NS free energy. We also recall that 
\be
F_{\rm NS}(t, \hbar) ={1\over \hbar} F_0(t) + \CO(\hbar), 
\ee
where $F_0(t)$ is the genus zero free energy of the toric CY in the large radius frame,
\be
F_0(t) ={C\over 6} t^3 + F_0^{\rm inst}(t). 
\ee

It is instructive to verify how the conventional Bohr--Sommerfeld quantization condition emerges from the exact quantization condition (\ref{eqc}). 
In the standard WKB limit (\ref{semi-lim}), the spectrum is of the form 
\be
E_n \approx E(t), 
\ee
where the function $E(t)$ is determined by the condition 
\be
\label{df-dual}
r \left({\partial F_0 \over \partial t}\right)_{t=t(E)}+ B = 2 \pi t. 
\ee
This is indeed of the form (\ref{bs}), and we learn in addition that
\be
\label{action-f0}
I(E)= {1\over 2 \pi}\left\{ r \left({\partial F_0 \over \partial t}\right)_{t=t(E)}+ B  \right\}. 
\ee

Let us now consider a non-interacting Fermi gas of $N$ particles in which the one-particle density matrix is $\rho$. One surprising result from 
\cite{ghm,mz} is that the limit in which one makes contact with the conventional topological string is not the standard WKB limit of the gas
(\ref{ds-largeN}), but rather the non-conventional 
limit (\ref{sc-limit}). As mentioned in the Introduction, it was conjectured in \cite{ghm} that, in this limit, 
the canonical partition function of the Fermi 
gas, $Z(N, \hbar)$, has the asymptotic expansion (\ref{as-zn}). In this expansion, 
$F_g(\lambda)$ is the genus $g$ topological string free energy of the CY $X$ in the so-called conifold frame, 
and $\lambda$ is a flat coordinate (in particular, it vanishes 
at the conifold point). Therefore, the all-genus topological string emerges in the limit (\ref{sc-limit})) of the Fermi gas, which provides a non-perturbative 
definition of the topological string partition function. 

In the following, we will be interested in analyzing the quantum theory in the limit (\ref{sc-limit}), which is the semiclassical limit of the topological string. In the 
quantum Fermi gas, we can regard it as a ``dual" semiclassical limit, in which the dual Planck constant 
\be
\hbar_D ={4 \pi^2\over \hbar}
\ee
goes to zero. We will now show that (\ref{sc-limit}) is 
effectively a low temperature limit for the non-interacting Fermi gas. To see this, let us study the spectrum of $\mO$ when
\be
\label{dual-sc}
\hbar_D \rightarrow 0, \qquad n \rightarrow \infty, \qquad \hbar_D n=\xi_D \, \, \text{fixed}.
\ee
The key fact to understand this regime is that, as emphasized in \cite{hatsuda-comments}, the exact quantization condition (\ref{eqc}) is invariant under the S-duality transformation
 \be
 \label{s-duality}
 t \rightarrow {2 \pi t \over \hbar}, \qquad \hbar \rightarrow \hbar_D.  
 \ee
This sort of invariance is expected from the modular duality of Weyl operators \cite{faddeev}. After multiplication by $4 \pi^2/\hbar^2$, (\ref{eqc}) can be written as 
 \be
 \label{eqc-dual}
 {C r \over 2} \left( {2 \pi t \over \hbar}\right)^2 + B(\hbar_D)+ \hbar_D\left( f_{\rm NS}\left( t, \hbar \right) + f_{\rm NS} \left( {2 \pi t \over \hbar}, \hbar_D \right) \right)=
  2 \pi \hbar_D \left( n+{1\over 2} \right), \qquad n=0,1,2, \cdots
  \ee
 From this form of the quantization condition, it is clear that, when $\hbar \rightarrow \infty$, 
 $t$ (and $E$) should scale like $\hbar$ (this scaling was already noted in \cite{hw}). Let us then assume that, in the limit (\ref{dual-sc}), the energy levels 
 behave like
 \be
 \label{en-dual}
 E_n \approx \hbar \CE (\xi_D), 
 \ee
and let us determine the function $\CE (\xi_D)$. Since $t,E\approx \hbar$ are large, 
we can drop exponentially small corrections in these quantities, like those appearing 
 e.g. in the quantum mirror map $t(E, \hbar)$. We also have
\be
\hbar_D f_{\rm NS}\left( {2 \pi t \over \hbar}, {4 \pi^2 \over \hbar} \right)= r {\partial F_0^{\rm inst} \over \partial t}\biggl|_{t\rightarrow {2\pi t \over \hbar}} + \CO(\hbar_D).
\ee
The equation determining $\CE$ as a function of $\xi_D$ is then
\be\label{lh-qc}
r  \left({\partial F_0 \over \partial t} \right)_{t= 2 \pi r \CE } + B= 2 \pi \xi_D. 
\ee

It is important to point out that, in spite of the formal invariance of the exact quantization condition under the transformation (\ref{s-duality}), the spectrum 
itself is {\it not} invariant. In fact, the spectrum scales like $\CO(\hbar^0)$ in the standard semiclassical limit $\hbar\rightarrow 0$, while it scales like $\CO(\hbar)$ 
in the dual limit $\hbar \rightarrow \infty$. However, the comparison between (\ref{lh-qc}) and (\ref{action-f0}) suggests introducing a ``dual" energy $ E_D$ through the equation 
\be
2 \pi r \CE =t(E_D), 
\ee
in such a way that the quantization condition (\ref{lh-qc}) reads now 
\be
\label{dbs}
I(E_D)= \xi_D
\ee
and it has the same form as (\ref{bs}). This dual energy will be important in order to describe the emergent classical geometry in the limit (\ref{sc-limit}). 

One consequence of (\ref{en-dual}) is that the limit (\ref{dual-sc}) is effectively a {\it zero temperature limit}, since in (\ref{en-dual}) the $\hbar$ factor acts like an effective inverse 
temperature $\beta=(k_BT)^{-1}$ for small $T$. In other words, in any thermal computation involving the Hamiltonian $\mH$ in (\ref{rhoH}), 
we can regard the limit (\ref{dual-sc}) as a limit in which we take simultaneously a zero-temperature limit and a WKB limit with effective energy levels 
given by $\CE(\xi_D)$. 
 
As a further check of this picture, we can compute the partition function $Z(N, \hbar)$ in the limit (\ref{sc-limit}) at leading order in $N^2$ (a similar 
calculation was performed in \cite{bipz}). Since we have a system of $N$ fermions at zero temperature, its canonical partition function is approximately given by 
 \be
 Z(N, \hbar) \approx \re^{-\CG}, 
 \ee
 where
 \be
 \CG=\sum_{n=0}^{N-1} E_n
 \ee
  is the energy of the ground state of the Fermi gas with $N$ particles. The energy levels $E_n$ in the limit (\ref{sc-limit}) are given by (\ref{en-dual}) with 
  \be
  \xi_D= 4 \pi^2 \zeta \lambda, \qquad  \zeta={n \over N}. 
  \ee
At large $N$, $\zeta$ can be regarded as a continuous parameter that varies between $0$ and $1$, and 
\be
\sum_{n=0}^{N-1} \rightarrow N \int_0^1 \rd \zeta. 
\ee
If we write
  \be
  \CG \approx -\hbar^2 {\mathcal F}_0(\lambda), 
  \ee
we find
  \be
  \label{f0}
  -{\mathcal F} _0(\lambda) \approx {1\over \hbar^2} \sum_{n=0}^{N-1} 
  E_n \approx {N \over 2 \pi r \hbar} \int_0^1 t \left(\lambda \zeta \right) \rd \zeta ={1\over 2 \pi r} \int_0^\lambda t(u) \rd u, 
  \ee
  where we changed variables to $u=\lambda \zeta$, and we wrote $t$ as the function of $\lambda$ defined implicitly by (\ref{lh-qc}), i.e. by 
  \be
  \label{lamder}
 r {\partial F_0 \over \partial t} + B = 8 \pi^3 \lambda. 
  \ee
By taking a derivative of (\ref{f0}) w.r.t. $\lambda$, we conclude that 
  \be
  \label{dert}
  {\partial {\mathcal F}_0 \over \partial \lambda}= -{t(\lambda) \over 2 \pi r}.  
  \ee
 The equations (\ref{lamder}), (\ref{dert}) are precisely the equations defining $\lambda$ as a conifold flat coordinate, and ${\mathcal F}_0 (\lambda)$ as the prepotential in the 
 conifold frame. They agree with the explicit calculations in 
 \cite{mz, kmz}. 
 
\begin{figure}
\begin{center}
	\includegraphics[width=0.7\textwidth]{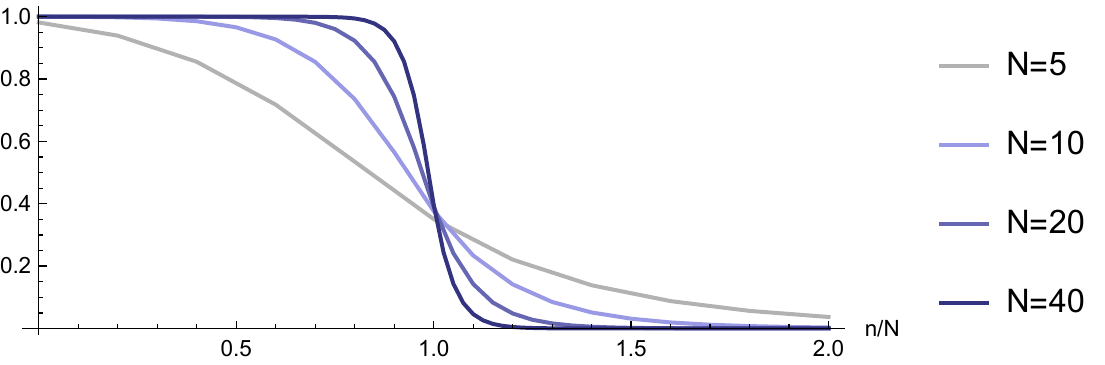} 
	\caption{
	The numerical values of $Nc_k^{(N)}$ as function of $k$, for different values of $N$. Here $\hbar$ is chosen so that $\lambda=N/\hbar=1$. As $N$ becomes large, 
	the coefficients display the step behavior (\ref{ckn-lowT}) typical of a Fermi gas at zero temperature.}
	\label{fig2}
\end{center}
\end{figure}

 As a further verification of the low-temperature nature of the limit (\ref{sc-limit}), let us look at a concrete geometry. Our main example in this paper will be 
 the toric CY known as local $\IF_0$. In this case, the function $\CO(\re^x, \re^p)$ appearing in (\ref{mirror-curve}) is given by 
 \be
 \label{mc-f0}
 \CO(\re^x, \re^p)=\re^p+\re^{-p} + \re^x+ \re^{-x}. 
 \ee
The reader should be aware that, for simplicity, we are setting to one the ``mass parameter" appearing in this geometry. The corresponding quantum operator is 
\be
\label{op-f0}
\mO=  \re^{\mm}+\re^{-\mm} + \re^{\mx}+ \re^{-\mx}. 
\ee
The spectral problem for this operator has been studied intensively in the last years, see for example 
\cite{km,hw,ghm,wzh,mz-wv,kaserg,butterfly,gm-complex,Sciarappa-2,gu-s,cms-np}. The energy levels of the 
corresponding Hamiltonian $\mH$ are given by the (conjectural) 
exact quantization condition (\ref{eqc}) with $r=2$, $C=1$, and $B=-2 \pi^2/3$. Using the spectrum of $\mH$ and the explicit expression (\ref{ckN}), 
it is possible to calculate numerically the 
coefficients $c_k^{(N)}(\hbar)$ appearing in (\ref{cn-ft}). We can then study their behavior in the limit (\ref{sc-limit}). The results for $N/\hbar=1$ and increasingly larger 
values of $N$ (and $\hbar$) are shown in \figref{fig2}. It is clear that in the large $N$, large $\hbar$ limit, they display the typical behavior (\ref{ckn-lowT}) 
of a non-interacting Fermi gas at zero temperature.

 % The above discussion, together with the exact S-duality structure for the eigenfunctions, suggests that in the strong coupling limit $\hbar\rightarrow \infty$ we have a 
 %system of $N$ fermions at essentially zero temperature, and where the ``classical" limit takes place in the `` dual" curve
 %
 %\be
% \CO (\widetilde x, \widetilde y) =\re^{\widetilde E},  
% \ee
 %
% where
 %
% \be
% \widetilde x= {2 \pi x  \over \hbar}, \qquad 
%  \widetilde y= {2 \pi y  \over \hbar}.
%  \ee
  %

%We should check that indeed this is the boundary of $C_N(q,p)$. 

\subsection{Reduced density matrix as a matrix model}
\label{sec-red-mm}

Let us consider the non-interacting Fermi gas of $N$ particles associated to a given mirror curve, which we will suppose of genus one to simplify our presentation. The density matrix for 
the one-body problem is defined by (\ref{rho-def}), but we would like to study the reduced density matrix for $N$ particles, (\ref{cn-gen}).  
In this section we will show that, in many examples, $C_N(x,y)$ can be written as a correlator in a matrix model. 
The limit (\ref{sc-limit}) turns out to be the 't Hooft limit of the matrix integral, and this makes it possible to obtain explicit 
expressions for $C_N(x,y)$ in this limit. The connection between the reduced density matrix of non-interacting 
fermions in a harmonic trap and the Gaussian matrix model has been useful in order to obtain analytic results for this system, as shown in e.g. \cite{dlms,entropy}. The results 
in this section can be regarded as a generalization of this connection to a wide 
class of Fermi systems associated to mirror curves.

It was found in \cite{kama} that, after an appropriate canonical transformation, the integral kernel of the operator $\rho$ for various mirror curves of genus one
 is of the form
\be
	\label{rhoGen1}
	\rho(\mu,\mu') =\frac{{v}(\mu)^{1/2} {v}(\mu')^{1/2} }{2 \cosh\left(\frac{\mu-\mu'}{2\gamma}+\ri \pi C \right)},
\ee
where $\gamma$ is real, $C$ is rational and $v(\mu)$ is a positive function which is bounded from above. Here, the variable $\mu$ is an appropriate combination
 of the original variables appearing in the mirror curve. The choice of $\mu$ is such that $\gamma$ does not depend on $\hbar$. The function $v(\mu)$ turns out to admit the representation
\be
	v(\mu) =\re^{-\hbar V(\mu)},
\ee
where
\be
\label{pot-exp}
	V(\mu) = V_{0}(\mu)+\CO(\hbar^{-2}).
\ee

 \begin{example} {\it Local $\IF_0$}. As an example of this structure, let us consider in some detail the case of local $\IF_0$. 
 The integral kernel of the operator $\rho$ was obtained in \cite{kmz}. It has a simple expression in the coordinates $q$, $P$, obtained from the coordinates $x$, $p$ appearing in (\ref{mc-f0}) by the following canonical transformation, 
%(\ref{
%
\be
\label{slc}
\ba 
x&={1\over {\sqrt{2}}} (q+P), \\
p&={1\over {\sqrt{2}}} (q-P). 
\ea
\ee
One finds, 
\be
	\label{rhoq1q2}
	\rho(q_1,q_2) = \frac{1}{\hbar\sqrt{2}}    \frac{ \re^{-\frac{\hbar}{2} W(q_1) -\frac{\hbar}{2} W(q_2) } }{2\cosh\left(\frac{\pi}{\hbar\sqrt{2}}(q_1-q_2) \right)},
\ee
where
\be
	W(q) = -\frac{1}{\hbar \sqrt{2}}q-\frac{2}{\hbar} \log \fad \left (\frac{1}{\hbar \mb} \left ( \frac{1}{\sqrt{2}}q+\frac{\ri \hbar}{4} \right ) \right )
	+\frac{2}{\hbar} \log \fad \left (\frac{1}{\hbar \mb} \left ( \frac{1}{\sqrt{2}}q-\frac{\ri \hbar}{4} \right ) \right ),
\ee
the parameter $\mb$ is given by 
\be
	\mb^2 = \frac{\hbar}{\pi}, 
\ee
and $\fad(x)$ is Faddeev's quantum dilogarithm. Let us now introduce the rescaled variable
\be
	\mu = \frac{\pi\sqrt{2}}{\hbar}q. 
\ee
The integral kernel (\ref{rhoq1q2}) reads, in these new variables, 
\be
\label{xik}
	\rho(\mu_1,\mu_2)= \rho(q_1(\mu_1),q_2(\mu_2))\frac{\rd q_2}{\rd \mu_2} = \frac{1}{2\pi}  \frac{ \re^{-\frac{\hbar}{2} V(\mu_1) -\frac{\hbar}{2} V(\mu_2) } }{2\cosh(\frac{1}{2}(\mu_1-\mu_2) )},
\ee
where
\be
	V(\mu) = -\frac{1}{2\pi}\mu+\frac{2}{\hbar} \log \Phi_b \left (\frac{\mu}{2\pi}-\frac{\ri}{4}  \right )
	-\frac{2}{\hbar} \log \Phi_b \left (\frac{\mu}{2\pi}+\frac{\ri}{4}  \right )
	\ee
has the expansion (\ref{pot-exp}) with 
\be
\label{vomu}
V_0(\mu)= -\frac{\mu}{2\pi}+\frac{2}{\pi^2}{\rm Im }\, {\rm Li}_2(\ri \re^\mu). 
\ee
This integral kernel (\ref{xik}) has indeed the form (\ref{rhoGen1}) with $\gamma=1$, $C=0$ (up to an overall normalization). \qed
\end{example} 

We compute the reduced density matrix from the expressions (\ref{cn-bn}) and (\ref{bn-integral}). By using the Cauchy determinant formula, as in 
\cite{kwy-ids,mp}, we get
\be
	\label{detCauchy1}
	{\det}_N \rho(\mu_i,\mu_j) = \prod_{i=1}^{N} v(\mu_i) \frac{ \prod_{i > j}\left ( 2\sinh \frac{\mu_i-\mu_j}{2\gamma} \right )^2}{ \prod_{i,j} 2 \cosh \left ( \frac{\mu_i-\mu_j}{2 \gamma}+\ri \pi C \right ) },
\ee
and
\be
			\rho
			 \begin{pmatrix}
				\mu & \mu_1 & ... & \mu_{N} \\
				\mu' & \mu_1 & ... & \mu_{N} \\
			\end{pmatrix}
			= \re^{-2\pi \ri C N} \rho(\mu ,\mu') \prod_{i=1}^N t_C \left ( \frac{\mu-\mu_i}{2\gamma} \right ) t_C \left ( \frac{\mu'-\mu_i}{2\gamma} \right )  {\det}_N \rho(\mu_i,\mu_j) ,
\ee
where
\be
	t_C(z) = \frac{\re^{\ri \pi C}\sinh(z)}{ \cosh(z+\ri \pi C) }.
\ee

Let us denote by 
\be
	\left \langle f(\mu_1,...\mu_N) \right \rangle = \frac{1}{Z_N}  \frac{1}{N!} \int_{\mathbb R^N} \rd \mu_1 \cdots \rd \mu_N   f(\mu_1,...\mu_N) {\det}_N \rho(\mu_i,\mu_j)
\ee
an expectation value in the matrix integral defined by (\ref{zn-int}). 
Then, by using  (\ref{bn-integral}) and (\ref{cn-bn}), we obtain the expression
\be
\ba
	& \frac{B_N(\mu,\mu')}{Z_N} = \re^{-2\pi \ri C N} \rho(\mu,\mu') \left \langle \prod_{i=1}^N t_C \left ( \frac{\mu-\mu_i}{2\gamma} \right ) t_C \left ( \frac{\mu'-\mu_i}{2\gamma} \right ) \right \rangle \\
		& =  \re^{-2\pi \ri C N} \rho(\mu,\mu') \, {\rm exp} \left [  \sum_{s=1}^\infty \frac{1}{s!}\left \langle \left ( \sum_{i=1}^N  \log t_C \left ( \frac{\mu-\mu_i}{2\gamma} \right ) +  \sum_{i=1}^N  \log t_C \left ( \frac{\mu'-\mu_i}{2\gamma} \right ) \right )^s \right  \rangle^{(c)} \right ],
\ea
\ee
where the superscript $(c)$ means that we use connected correlators in the matrix model. Let us introduce the exponentiated variable $M=\re^{\mu/\gamma}$, 
and the function
\be
	W(M) = \sum_{i=1}^N \log t_C \left ( \frac{\mu-\mu_i}{2\gamma} \right ).
\ee
We can then write
\be
\ba
	\frac{B_N(\mu,\mu')}{Z_N} &= \re^{-2\pi \ri C N}  \rho(\mu,\mu') \, {\rm exp} \left [  \sum_{s=1}^\infty \sum_{\ell=0}^s \frac{1}{(s-\ell)! \ell!}\left \langle W(M)^{s-\ell} W(M')^\ell \right  \rangle^{(c)} \right ] \\
	 &= \re^{-2\pi \ri C N}  \rho(\mu,\mu') \, {\rm exp} \left [ 
	 \sum_{n=1}^\infty \frac{1}{n!}  \left \langle W(M)^n \right  \rangle^{(c)} 
	 +\sum_{n=1}^\infty \frac{1}{n!}  \left \langle W(M')^n \right  \rangle^{(c)}  \right. \\
	 & \qquad \qquad \qquad \qquad \qquad \qquad  \left.
	 + \sum_{m,n=1}^\infty  \frac{1}{m! n !}\left \langle W(M)^{m} W(M')^n \right  \rangle^{(c)} \right ]. 
\ea
\ee
Similar expressions appear in the study of annulus amplitudes in non-critical string theory, see e.g. \cite{kopss}. We now define the $n$-point function as
\be
\ba
	W_{n}(M_1, ... ,M_n)
	&= \frac{\partial}{\partial M_1} \cdots \frac{\partial}{\partial M_n} \left \langle W(M_1) \cdots W(M_n) \right  \rangle^{(c)} \\
	&=
	 \left \langle \prod_{k=1}^n \sum_{i=1}^N  \left ( \frac{1}{M_k-\re^{\mu_i/\gamma}} -  \frac{ 1}{ M_k-\omega \re^{\mu_i/\gamma}} \right ) \right  \rangle^{(c)},
\ea
\ee
where we set
\be
	\omega =-\re^{-2\pi \ri C}.
\ee
This is the analogue in this case of the $n$-point function for Hermitian matrix integrals (see e.g. \cite{ackm})
\be
	W_{n}(M_1, ... ,M_n) = 
	 \left \langle \prod_{k=1}^n \sum_{i=1}^N  \frac{1}{M_k-\mu_i} \right  \rangle^{(c)}.
\ee
As in the Hermitian case, we expect that, in the 't Hooft limit (\ref{sc-limit}), these $n$-point functions have an asymptotic expansion of the form
\be
	W_{n}(M_1, ... ,M_n) = \sum_{g =0}^{\infty} \hbar^{2-2g-n} W_{n,g}(M_1,...,M_n).
\ee
We finally obtain, to next-to-leading order in the large $N$ expansion, 
\be
	\label{BNoZNanalytic}
\ba
	& \frac{B_N(\mu,\mu')}{Z_N} =\re^{-2\pi \ri C N} \rho(\mu,\mu') \, {\rm exp} \left [  \langle W(M) \rangle^{(c)} + \langle W(M') \rangle^{(c)}  \right. \\
	& \qquad \qquad \qquad  \qquad \left. + \frac{1}{2} \langle W(M)^2 \rangle^{(c)} + \frac{1}{2} \langle W(M')^2 \rangle^{(c)} + \langle W(M) W(M') \rangle^{(c)}+\CO(N^{-1}) \right ] \\
	&=  \rho(\mu,\mu') \, {\rm exp} \left [ -2\pi \ri C \lambda \hbar+ 
		\hbar  \int_\infty^M W_{1,0}(Z)\rd Z +\hbar  \int_\infty^{M'} W_{1,0}(Z)\rd Z \right. \\
	& \qquad \qquad \qquad 
		+\frac{1}{2} \int_\infty^M \int_\infty^M W_{2,0}(Z_1,Z_2)\rd Z_1 \rd Z_2  +
		\frac{1}{2} \int_\infty^{M'} \int_\infty^{M'} W_{2,0}(Z_1,Z_2)\rd Z_1 \rd Z_2 \\
	& \qquad \qquad \qquad \left.
		+ \int_\infty^{M} \int_\infty^{M'} W_{2,0}(Z_1,Z_2)\rd Z_1 \rd Z_2+ \CO(\hbar^{-1})\right ].
\ea
\ee 
By using (\ref{cn-bn}), we can obtain from the above expression the reduced density matrix $C_{N+1}(x,y)$ for many Fermi gases associated to 
quantum mirror curves, as well as its behavior in the semiclassical limit. 
The expression (\ref{BNoZNanalytic}) involves standard one and 
two-point correlation functions of the matrix model associated to the Fermi gas. 
As explained in \cite{kmz}, when $C=0$ the resulting matrix model can be mapped to an $O(2)$ matrix model, and the functions $W_{1,0}(M)$, $W_{2,0}(M,M')$ 
can be explicitly calculated from the results in \cite{ek1,ek2}. In Appendix \ref{app-localF0} we present such a calculation in the case of local $\IF_0$. 

\subsection{Quantum geometry and its classical limit}
\label{qg-cl}

Given a genus one mirror curve, we have considered the non-interacting, $N$ particle Fermi gas associated to it, and in particular its reduced density matrix. 
We can also consider the quantum distribution (\ref{wigner-reduced}), which is a function on phase space depending on $N$ and $\hbar$. We will be interested in this distribution in the limit (\ref{sc-limit}). In this limit, the Fermi energy 
scales as $\hbar$, as we explained in section \ref{sec-ts-sl}, and the region in phase space where $\CW_N(x,p)$ is non-negligible grows with $\hbar$. 
In order to have a ``stable" limit as $\hbar$ becomes larger, it is convenient to rescale the phase space variables. This is also suggested by the study of the open string 
wavefunction for the quantum mirror curve, which requires such a scaling of the position space coordinate in the limit (\ref{sc-limit}) \cite{mz-wv}. 
Therefore, we define the rescaled quantum distribution in phase space associated to a mirror curve as 
\be
\label{def-qd}
\CQ_N (x, p) = \left( {\hbar \over 2 \pi} \right)^2 \CW_N \left( {\hbar x \over 2 \pi}, {\hbar p \over 2 \pi}\right). 
\ee
This involves the phase space coordinates appearing in the modular double theory \cite{faddeev}. The prefactor guarantees that the rescaled 
distribution is correctly normalized. 

The quantum distribution (\ref{def-qd}) is an appropriate and precise definition of quantum geometry in the context of mirror curves. Indeed, we claim that in the 
limit (\ref{sc-limit}), this distribution has constant support in the interior of the mirror curve
\be
\label{dmc}
\CO(\re^x, \re^p) =\re^{E_{DF}}. 
\ee
Here, $E_{DF}$ is the dual Fermi energy, and it is determined by the 't Hooft parameter $\lambda$ through the equation 
\be
 \label{ed-lam}
 I(E_{DF})= 4 \pi^2  \lambda. 
 \ee
 This follows from (\ref{dbs}) with $\xi_D= 4 \pi^2 \lambda$. More precisely, we claim that, in the limit (\ref{sc-limit}), 
\be
\label{qn-limit}
\CQ_N (x, p)  \approx {1\over 2 \pi I(E_{DF})} \Theta\left( \re^{E_{DF}} - \CO(\re^x, \re^p) \right). 
\ee
Therefore, in this limit, the boundary of the support of $\CQ_N(x, p)$ is the classical mirror curve. 
Away from the limit (\ref{sc-limit}), the quantum distribution $\CQ_N(x, p)$ exhibits 
fluctuations that make this boundary ``fuzzy" in a precise, quantitative way. 

As an illustrative example, let us consider again the local $\IF_0$ geometry. Using the explicit expression for the integral kernel (\ref{rhoq1q2}), it is in principle possible to 
compute analytically the reduced density matrix for rational values of $\hbar/\pi$ and low values of $N$, by using (\ref{cn-bn}), (\ref{bn-integral}), and the 
integration techniques developed in \cite{garou-kas}. E.g. for $N=1$ 
and $\hbar=2 \pi$ we find, 
\be
\label{n1}
\CQ_1(x, p)= {1\over 2 \pi} {1\over \cosh\left( {x-p \over 2} \right)} \left\{ 
{1\over 2 \cosh \left( {x+p \over 2} \right)}- {\sin \left( {x^2-p^2 \over 2 \pi} \right) \over \sinh \left( {x-p \over 2} \right) 
\sinh(x+p)} \right\}. 
\ee
In deriving this result we have also used that, under a linear canonical transformation like (\ref{slc}), 
the Wigner function transforms by a change of coordinates (this follows e.g. from the 
general results in \cite{cfz}). 

For a systematic study of the functions $\CQ_N(x, p)$ for higher values of $N$, we use instead the expression (\ref{cn-ft}). After a 
Wigner transform, this involves the Wigner functions of the eigenstates of the operator $\rho$. Although Berry's formula (\ref{berry-f}) was originally derived for 
Hamiltonians of the form $H(x,p)= p^2/2+ V(x)$, we have explicitly verified that it correctly describes the Wigner 
functions associated to eigenfunctions of $\rho$. We will use however a more precise, numerical determination of these functions, 
obtained as follows. Let $\psi_n(x)$ denote the eigenfunctions for the harmonic oscillator, as in (\ref{eigen-ho}). It is well-known that the 
mixed Wigner functions associated to this basis can be computed 
in terms of generalized Laguerre polynomials \cite{b-moyal}. We have, 
\be
\ba
	E_{mn}(x,p) &\equiv  \frac{1}{2\pi \hbar} \int_{-\infty}^{\infty} \psi_{m}^* \left (x+\frac{y}{2} \right )\psi_{n} \left (x-\frac{y}{2} \right ) \re^{\frac{\ri}{\hbar} p y} \rd y\\
	&=  \frac{1}{2\pi \hbar} \frac{2^{\frac{n+m}{2}+1}}{\sqrt{n! m!}} \re^{- z \bar z} \sum_{\ell=0}^{{\rm min}(m,n)} {m \choose \ell} {n \choose \ell} (-2)^{-\ell}\ell! \, z^{m-\ell} \, {\bar z}^{n-\ell},
\ea
\ee
where
\be
	z= x+\frac{\ri}{\hbar}p, \qquad \bar z= x-\frac{\ri }{\hbar}p.
\ee
We can now expand the eigenfunctions appearing in (\ref{cn-ft}) in the basis (\ref{eigen-ho}):
\be
	\varphi_n(x)=\sum_{i\ge 0} v_{in} \psi_{i}(x), \qquad \qquad v_{in} \in \mathbb C.
\ee
Then, the Wigner transform becomes
\be
	f_{n}(x,p) = \sum_{i,j \ge 0} v_{in} ^{*}v_{jn} E_{ij}(x,p) = (v^{\dagger} E v)_{nn}.
\ee
In a numerical calculation of the eigenfunctions $\varphi_n(x)$ by the Rayleigh--Ritz method, we determine approximate values of the coefficients $v_{ni}$ for $i=0, 1, \cdots, n_{\rm max}$, and this gives an approximation to the Wigner function $f_n(x,p)$. Our numerical calculations of (\ref{cn-ft}) involve a double truncation in the index $i$ and in the index $n$. 

 Let us now compare the quantum distribution with the limiting classical geometry. The classical action $I(E)$ can be found by using (\ref{action-f0}) and standard results in the 
 special geometry of local $\IF_0$ (see e.g. \cite{kmz}). It reads, 
\be
\label{ie-kappa}
	I(E)= \frac{\kappa}{4\pi^2} G_{3,3}^{2,3}\left(
\begin{array}{c}
 \frac{1}{2},\frac{1}{2},\frac{1}{2} \\
 0,0, -\frac{1}{2} \\
\end{array}
\left | \frac{\kappa^2}{16} \right.
\right) -\pi, 
\ee
where $E$ is related to $\kappa$ through (\ref{kappa-E}). 
The relationship between the 't Hooft parameter $\lambda$ and the modulus of the limiting curve is given by (\ref{ed-lam}). 

\begin{figure}[h!]
\begin{center}
	\vspace{0.1cm}
	\includegraphics[width=0.32\textwidth]{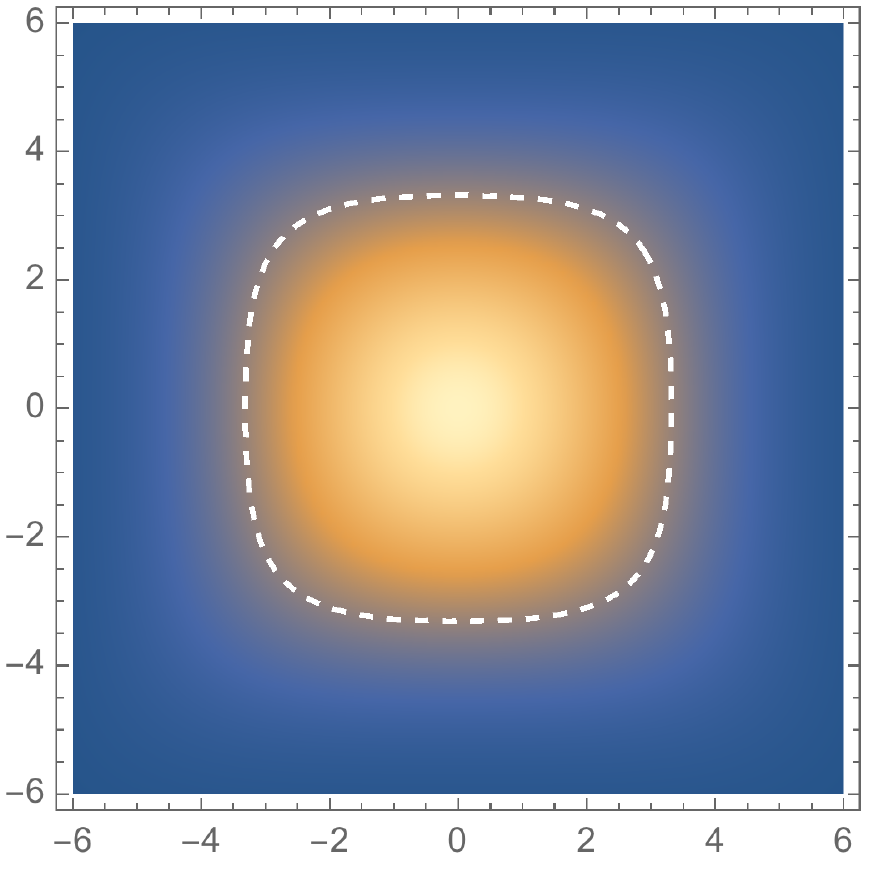}
	\includegraphics[width=0.32\textwidth]{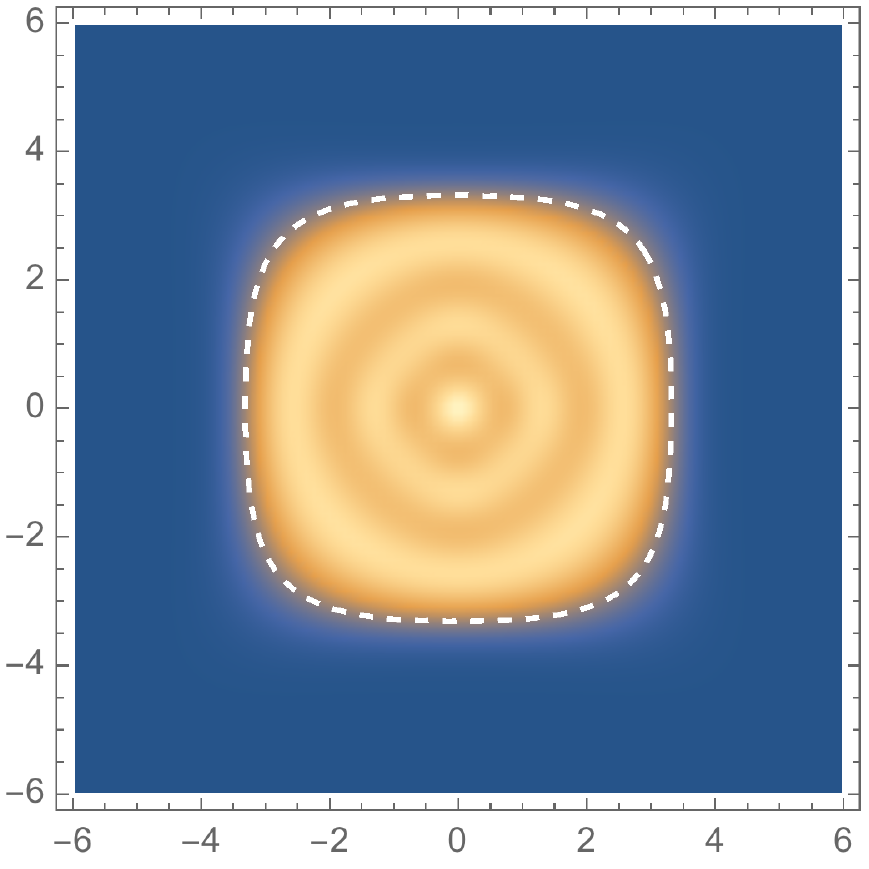}
	\includegraphics[width=0.32\textwidth]{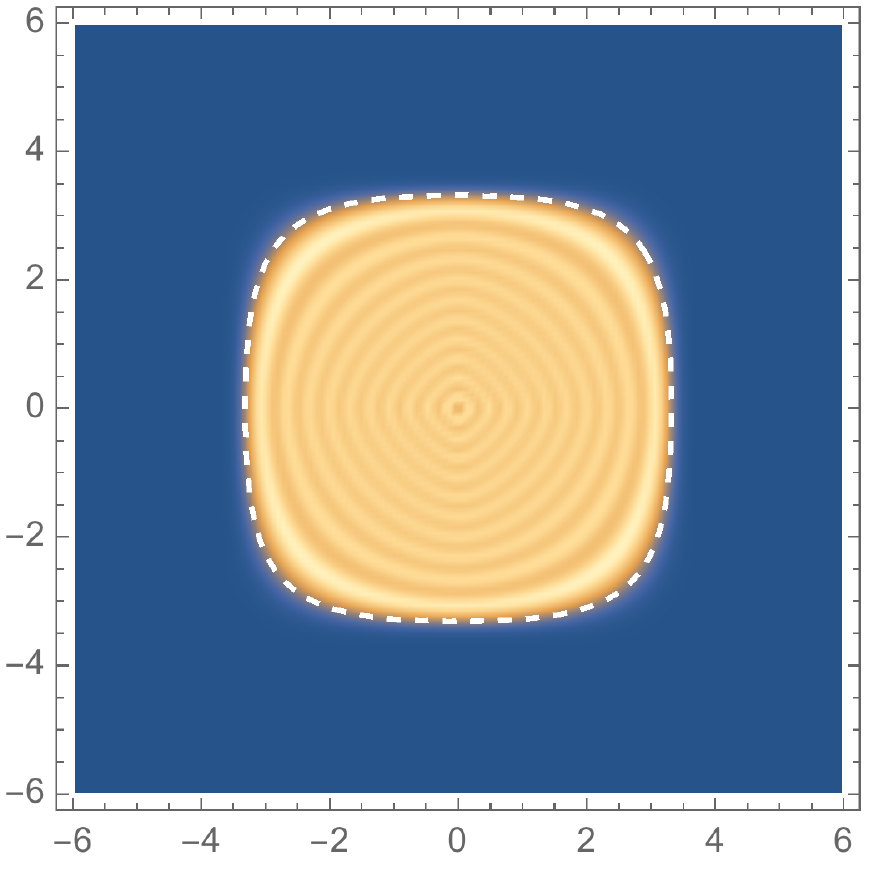} \\
	\hspace{0.25cm}
	\includegraphics[width=0.32\textwidth]{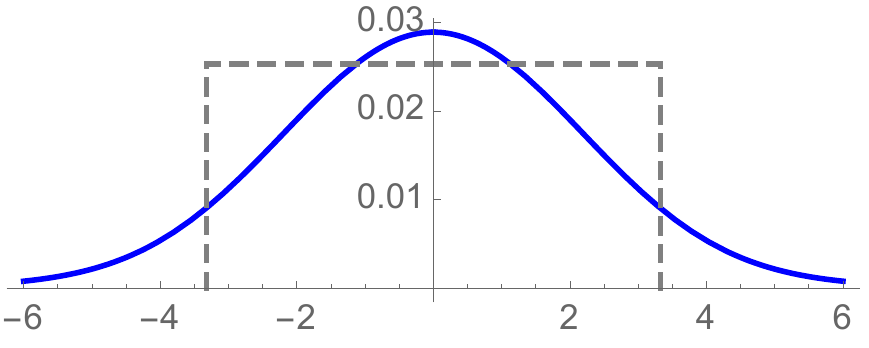}
	\includegraphics[width=0.32\textwidth]{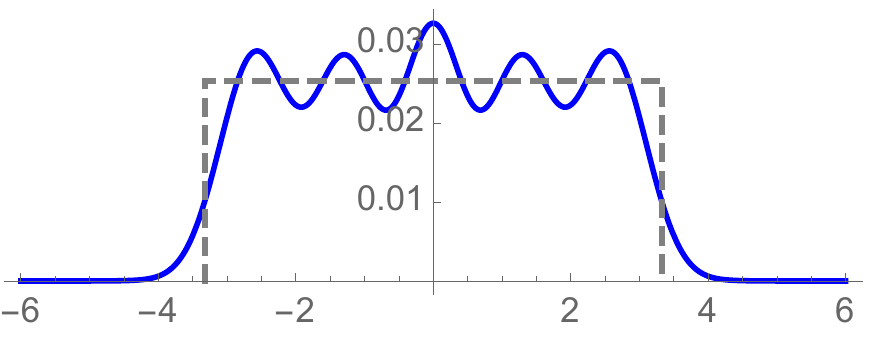}
	\includegraphics[width=0.32\textwidth]{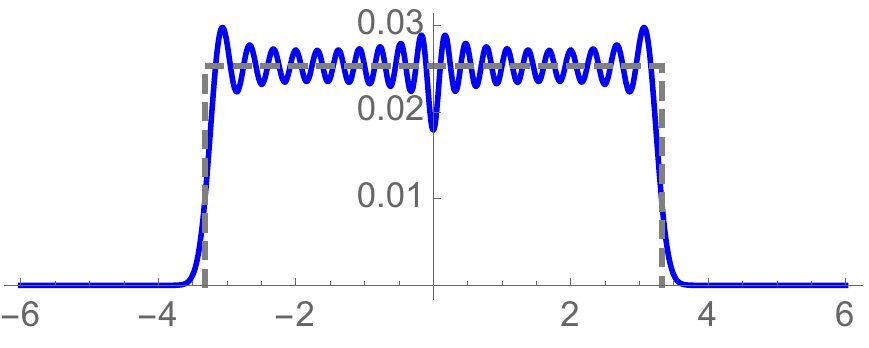}
		\caption{Top: Density plot of the quantum distribution $\CQ_N(x,p)$ in phase space for local $\IF_0$, compared to the classical mirror curve (\ref{dmc}), which is 
		shown as a white dashed line. In all cases, we have $N/\hbar=(2\pi)^{-1}$. On the left we show the distribution for $N=1$ (where the distribution is given by (\ref{n1})), in the middle we show 
		$N=5$, and on the right we show $N=20$. In the bottom, we show the restrictions of the same distributions to the slice $p=0$, as well the classical limit (\ref{qn-limit}) in dashed lines. }
		\label{comp-1}
\end{center}
\end{figure}

\begin{figure}[h!]
\begin{center}
		\vspace{1cm}
	\includegraphics[width=0.32\textwidth]{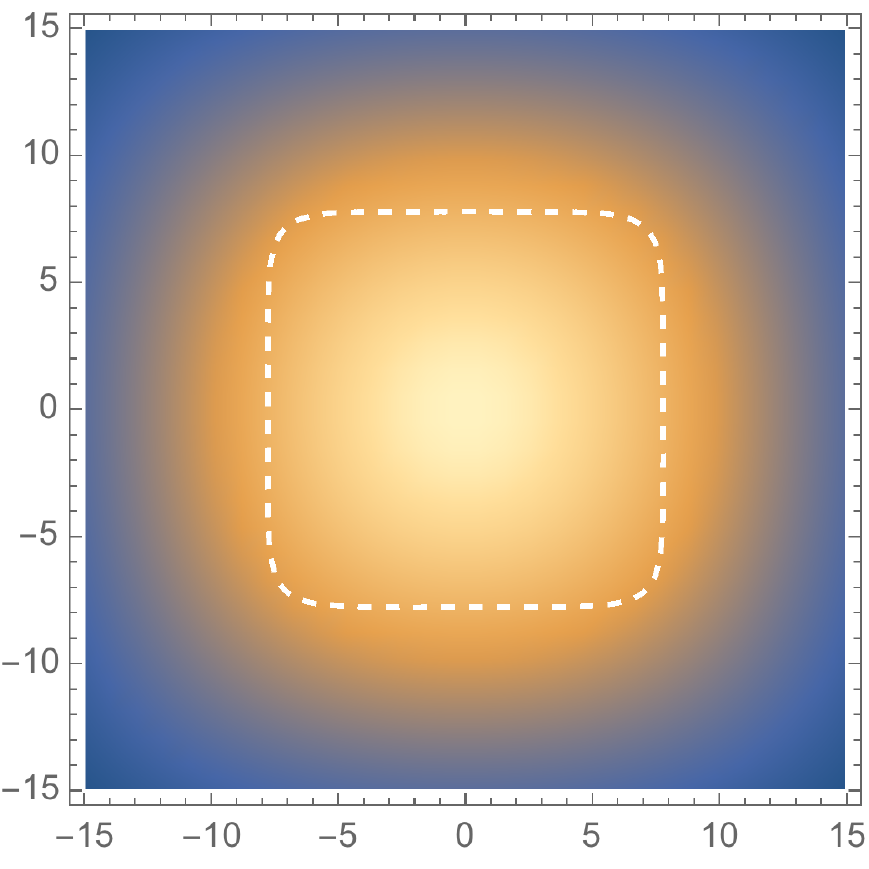}
	\includegraphics[width=0.32\textwidth]{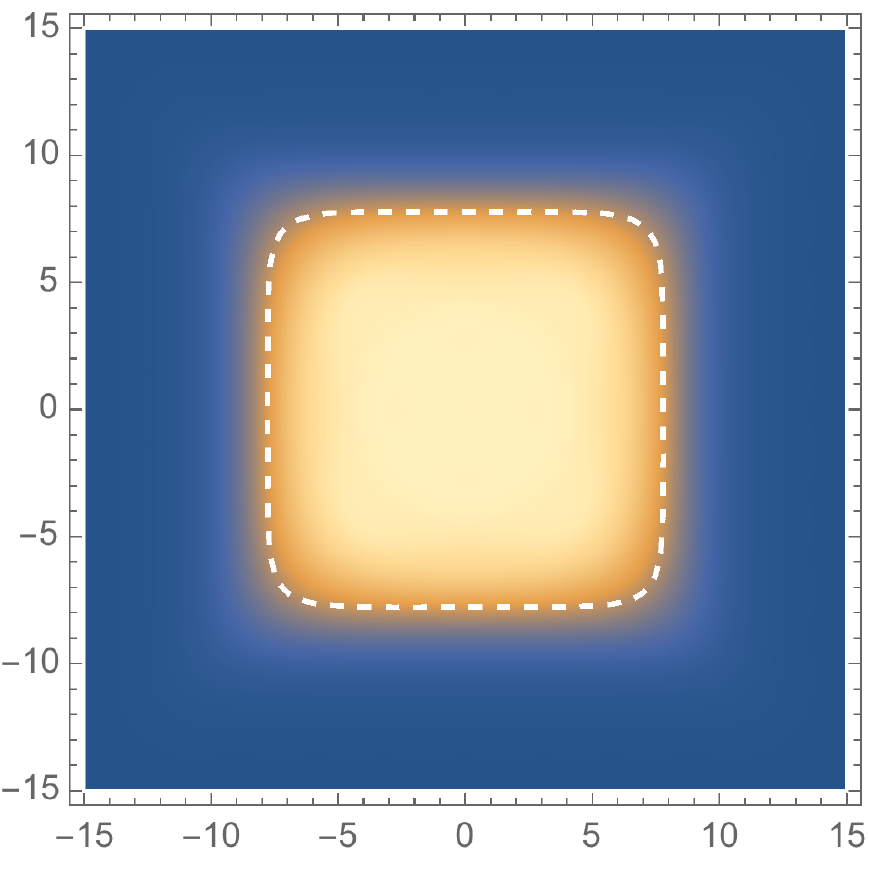}
	\includegraphics[width=0.32\textwidth]{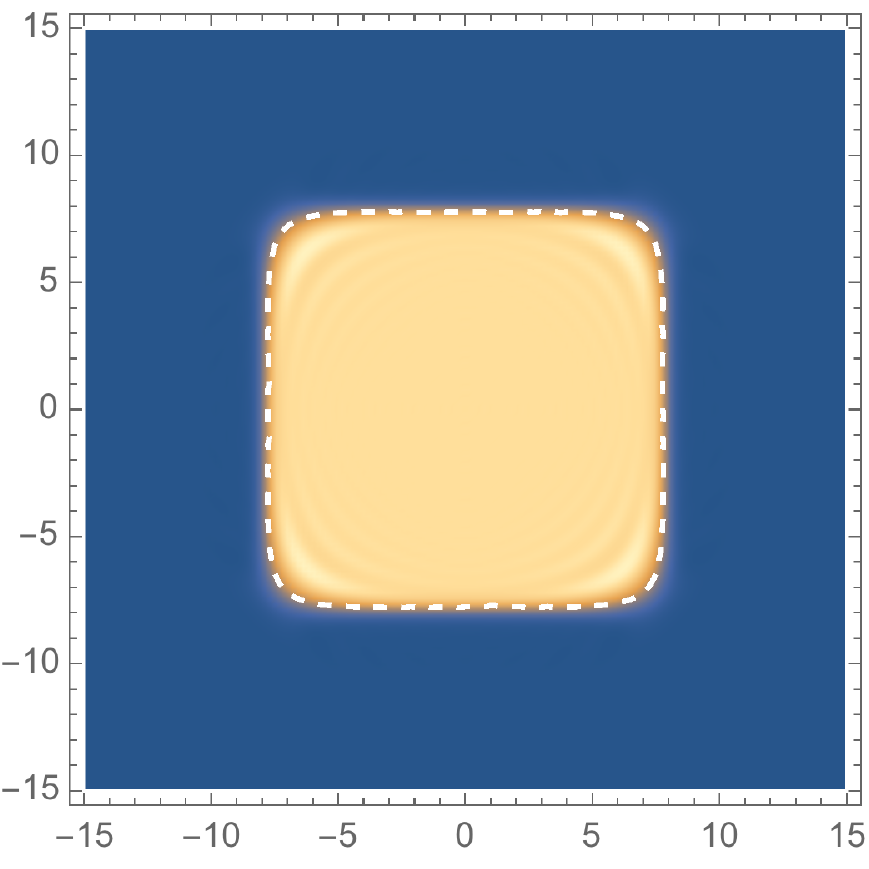} \\
	\hspace{0.25cm}
	\includegraphics[width=0.32\textwidth]{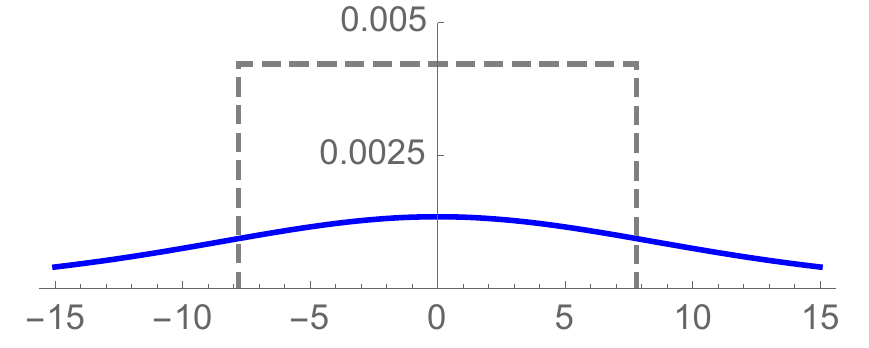}
	\includegraphics[width=0.32\textwidth]{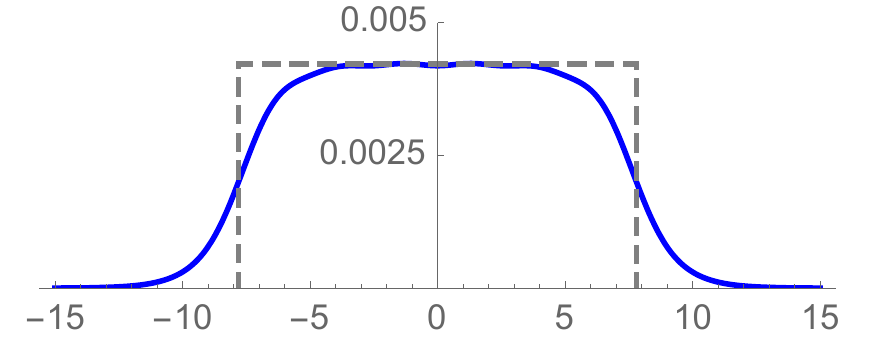}
	\includegraphics[width=0.32\textwidth]{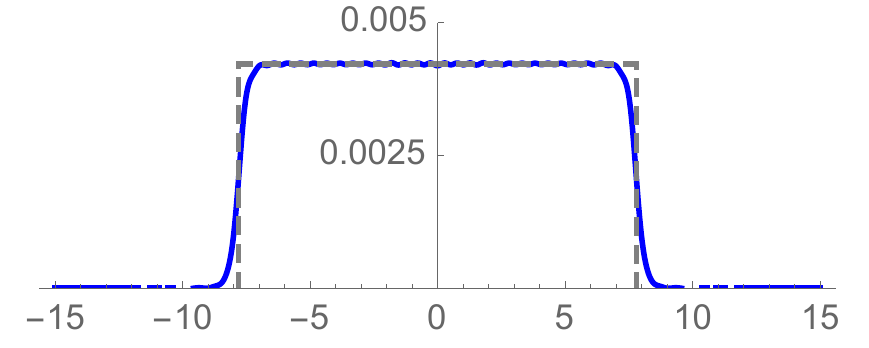}
		\caption{Same plots for $N/\hbar=3/\pi$. On the left $N=1$, in the middle $N=6$, on the right $N=30$.}
		\label{comp-2}
\end{center}
\end{figure}

In \figref{comp-1} and \figref{comp-2}, we compare the distribution $\CQ_N(x,p)$ to the expected limiting behavior in (\ref{qn-limit}), for various (increasing) 
values of $N$, and two fixed values of $\lambda$. It is clear that, as $N$ increases, the quantum distribution 
is more and more localized in the interior of the 
classical curve. Note that, from the point of view of quantum geometry, $N=1$ corresponds to a very quantum regime, 
in which the quantum distribution is spread out in a wide region around 
the classical limit. The emergence of the mirror curve as a sharp boundary, as $N$, $\hbar$ increase, can be seen very clearly in the density plots of the 
quantum distribution shown in \figref{comp-1} and \figref{comp-2}.

%\newpage 

% \newpage

\begin{figure}[h!]
\begin{center}
	\includegraphics[width=0.35\textwidth]{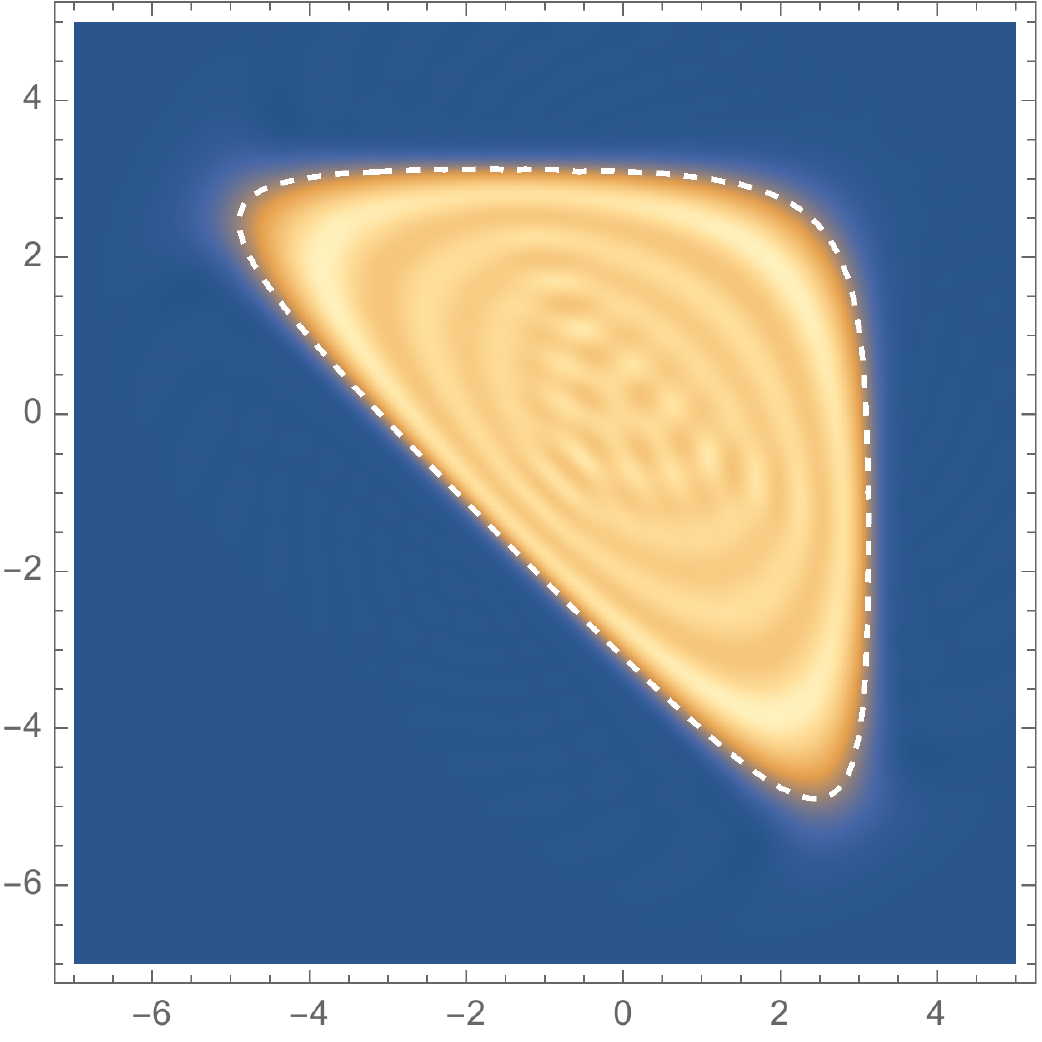} \\
	\hspace{0.25cm}
	\includegraphics[width=0.5\textwidth]{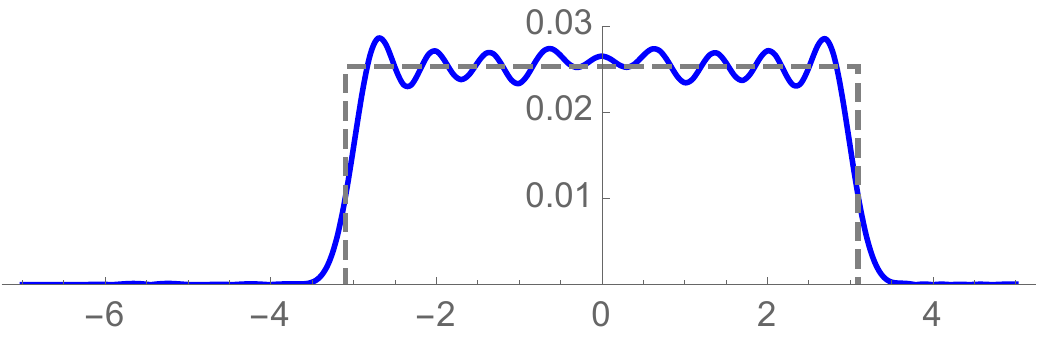} 
		\caption{The same plot for local $\mathbb P^2$ for $N/\hbar=1/2\pi$. Here, $N=10$. On the bottom plot, we show the restrictions of the Wigner distribution on the slice $p=0$, together with the classical limit in dashed.
		\label{figP2}}
\end{center}
\end{figure}

Although we have focused so far on the local $\IF_0$ geometry, we can consider other geometries of genus one, like local $\mathbb P^2$. In this case the mirror curve (\ref{mirror-curve}) 
corresponds to
\be
	\CO(x, p)=\re^{x} + \re^{p} + \re^{-x- p}, 
\ee
and the classical action is given by 
\be
\ba
	I(E) &= \frac{\sqrt{3} \Gamma \left( \frac{1}{3} \right )^3}{4\pi^2} \kappa \,\,_3 F_2 \left ( \frac{1}{3},\frac{1}{3},\frac{1}{3} ;\, \frac{2}{3},\frac{4}{3};\, \frac{\kappa^3}{27} \right )\\
	& -\frac{\sqrt{3} \Gamma \left( \frac{2}{3} \right )^3}{8\pi^2} \kappa^2 \,\,_3 F_2 \left ( \frac{2}{3},\frac{2}{3},\frac{2}{3} ;\, \frac{4}{3},\frac{5}{3};\, \frac{\kappa^3}{27} \right )-\frac{2\pi}{3},
	\ea
\ee
where $\kappa$ is as in (\ref{kappa-E}) the exponentiated energy. As we show in \figref{figP2}, where we compare the two sides of (\ref{qn-limit}) for local $\IP^2$, we also find in this example 
that the quantum distribution $\CQ_N (x,p)$ sharpens around the classical mirror curve in the limit (\ref{sc-limit}).

We would like to have as well a precise description of the quantum fluctuations of the distribution $\CQ_N ( x,  p)$ away from the 
strict classical limit (\ref{qn-limit}). It is natural to expect that 
the improved Balazs--Zipfel form (\ref{ibz}), suitably adapted to mirror curves, provides such a description, so that 
\be
	\label{bzFunc}
\CQ_N ( x,  p) \approx  \frac{1}{2\pi I(E_{DF})} {\mathcal I} \left ( - \left ( \frac{3 \mathcal A(x,p)}{2\hbar_D} \right )^{2/3} \right), 
\ee
where $\mathcal A(x,p)$ is the area of the chord defined by the classical mirror curve (\ref{dmc}). The above approximation can be derived analytically for the local $\IF_0$ geometry, as we show in Appendix \ref{app-localF0}, but we expect this scaling form to be valid for any genus one mirror curve. We can also 
obtain a transitional approximation near the classical curve, as in (\ref{ibz-trans}), which gives 
\be
 	\label{bzFunc-trans}
\CQ_N ( x,  p) \approx  \frac{1}{2\pi I(E_{DF})} {\mathcal I} \left ( 2 {I(x,p)-I(E_{DF})) \over \hbar^{2/3} B^{1/3}(x,p)}\right).
\ee

\begin{figure}[h]
\begin{center}
	\includegraphics[width=0.95\textwidth]{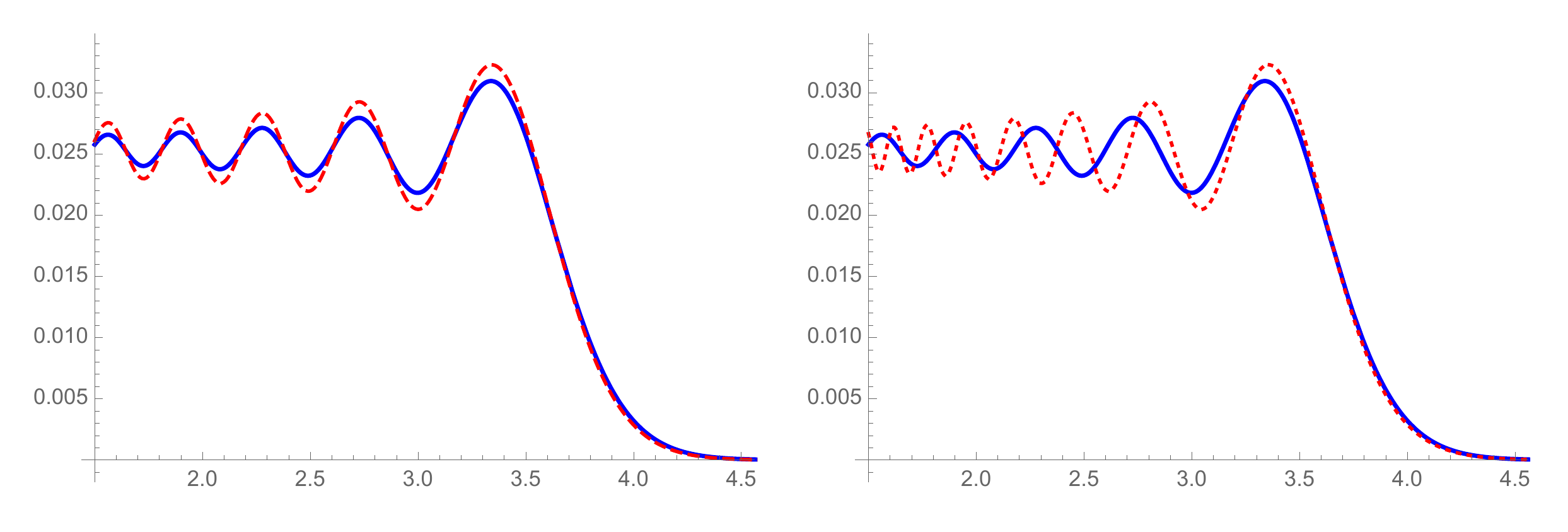} 
		\caption{Comparison of $\CQ_{N} (x,p)$ (in blue solid) with its semiclassical approximations, for local $\IF_0$. On the left, the red dashed line is the expression (\ref{bzFunc}) involving the chord area, which we analytically continue outside the classically allowed region. On the right, it is the transitional expression (\ref{bzFunc-trans}). We look at the slice $x=p$, and we set $N=20$, $\hbar=40 \pi$. }
\label{figBZnew2}
\end{center}
\end{figure}

In  \figref{figBZnew2} we compare the exact result for $\CQ_N(x,p)$, for local $\IF_0$, 
with the improved Balazs--Zipfel form in the r.h.s. of (\ref{bzFunc}), and with the transitional approximation (\ref{bzFunc-trans}), for $N=20$ and 
$\hbar=40 \pi$, so that $\lambda=(2 \pi)^{-1}$. As we can see, the expression (\ref{bzFunc}) involving the full chord area give a very good approximation deep inside the classical region. Both approximations 
capture with precision the shape of the quantum distribution in the vicinity of the classical mirror curve.

\sectiono{Conclusions and outlook}

In this paper we have shown that, given the mirror curve to a toric Calabi--Yau manifold, 
one can naturally define a quantum distribution on a two-dimensional phase space. To define this 
distribution, we first obtained a one-body density matrix by quantizing the mirror curve as in \cite{ghm}, and we then considered the 
corresponding non-interacting Fermi gas of $N$ particles. 
The sought-for distribution is just the Wigner transform of the reduced density matrix of this Fermi gas. We have argued 
that this distribution provides an appropriate definition of the quantum version 
of the mirror curve. This definition is non-perturbative in $\hbar$, or equivalently, in the string coupling constant. The 
semiclassical limit of the theory is the 't Hooft-like limit (\ref{sc-limit}), which corresponds to weakly interacting topological strings propagating on a CY 
background with a modulus determined by the 't Hooft coupling $\lambda$. 
We have conjectured that, in this limit, the quantum distribution becomes constant in the interior of the classical mirror curve, 
and vanishes outside. In other words, the classical 
mirror curve emerges in the semiclassical limit, as the boundary of the support of the quantum 
distribution. We have given numerical 
and analytical evidence for our conjecture, focusing on the example of local $\IF_0$. In addition, small fluctuations of the distribution 
around the limiting shape are captured 
by the universal scaling form (\ref{ibz}), which depends only on classical data of the mirror curve.

There are various questions opened by our investigation. First of all, our main claims, concerning the 
limiting shapes of the quantum distribution in the semiclassical regime, are conjectural. We have 
proved our claims in the case of local $\IF_0$ by a detailed calculation, but there might be a simpler and more general 
argument which establishes the conjectures on the limiting behavior for general genus one mirror curves. It would be also interesting to extend our results to mirror curves of 
genus $g_\Sigma$ larger than one. From the result of \cite{cgm}, we expect that this generalization will 
involve a Fermi gas with $g_\Sigma$ different types of particles. 

In our definition of the quantum curve, we have used the Wigner distribution 
associated to the reduced density matrix of the gas, but one could consider instead the Husimi distribution. In the semiclassical 
limit, this distribution also localizes on the classical curve in phase space, but it displays a Gaussian decay around it instead of an oscillatory behavior 
\cite{voros-husimi, saraceno}. 
For this reason, the Husimi distribution has been advocated in different contexts \cite{anderson, babel} as a more suitable definition of quantum geometry. 
It would be very interesting to work out the behavior of the Husimi distribution in the case of quantum mirror curves, both numerically and analytically.  

One should also explore in more detail 
the dictionary relating the physical properties of the Fermi gas to the underlying geometry of topological strings. For example, 
it is possible in some cases to calculate analytically the entanglement entropy of non-interacting Fermi gases (see e.g. \cite{entropy}). It would be 
interesting to see what is the geometric and physical 
counterpart of this quantity in the topological string side (a comparison along these lines for the $c=1$ string and its dual description has been made in \cite{hartnoll}). 

We have also found that the Balazs--Zipfel universal scaling function 
obtained in \cite{bzipfel} (which has been recently generalized to higher dimensions in \cite{dlms2}) 
can be slightly improved by using Berry's chord construction in \cite{berry}. It is tempting to conjecture that the improved 
Balazs--Zipfel approximation (\ref{ibz}) provides a better description 
of the Wigner distribution in the semiclassical regime, at least for one-dimensional systems 
(in fact, the transitional approximation (\ref{ibz-trans}) should already lead to an 
improvement of the scaling form near the classical curve). More generally, we believe that a 
precise scaling theory for Wigner distributions near the classical limit is still lacking. One should find 
appropriate scaling variables and a corresponding scaling limit in which the universal forms obtained by Berry and 
Balazs--Zipfel provide an exact description. A precise scaling theory can be obtained in the case 
of the harmonic oscillator, in which the Wigner distribution depends effectively on one single variable, namely the classical Hamiltonian. However, in the general 
case (even for one-dimensional 
problems), this theory is yet to be developed. 

\section*{Acknowledgements}
We would like to thank Santiago Codesido and Joan Sim\'on for useful discussions. This work is is supported in part by the Fonds National Suisse, 
subsidy 200020-175539, and by the NCCR 51NF40-141869 ``The Mathematics of Physics'' (SwissMAP).

\appendix 

\sectiono{Semiclassical distribution for the harmonic oscillator}
\label{ho} 

\begin{figure}[h]
\begin{center}
	\includegraphics[width=0.5 \textwidth]{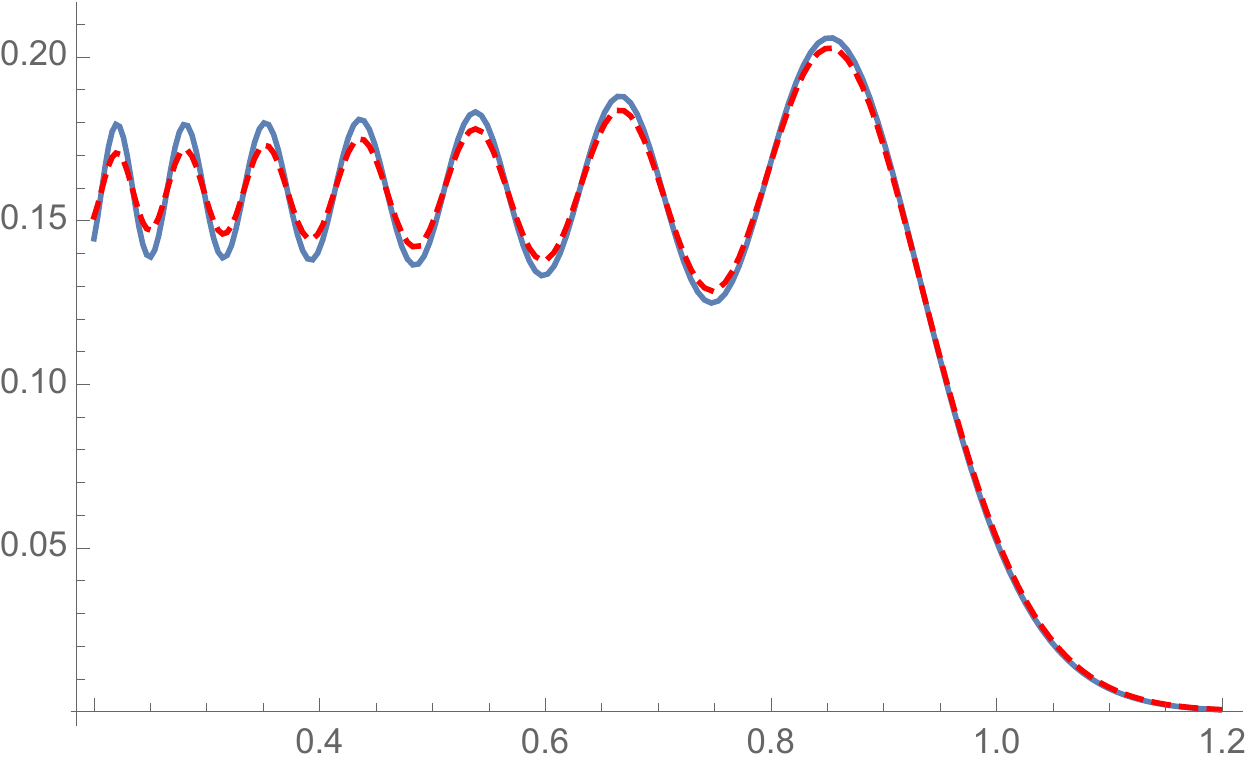}
	\caption{The improved Balazs--Zipfel approximation to the quantum distribution of non-interacting fermions at zero temperature in a harmonic potential, 
	(\ref{finalwigCNqho}) (dashed line), compared to the exact function $\mathcal W_{N}$ (full line), as a function of 
	$H(x,p)$, for $N=30$, $\hbar=1/30$, and $\xi=1$.}
	\label{ibz-ho}
\end{center}
\end{figure}

In this Appendix we derive the improved Balazs--Zipfel approximation for the Wigner transform of the reduced density matrix of a Fermi gas at zero 
temperature. We choose units in which $m=\omega=1$. By using (\ref{cn-ho}), we find
 \be
 	{\mathcal W}_{N}(x,p)  = \frac{1}{2\pi \sqrt{2 N \hbar}} \int_{\mathbb R} \rd y \, 
		\frac{ \psi_{N}\left (x+\frac{y}{2} \right ) \psi_{N-1}\left (x-\frac{y}{2} \right ) -\psi_{N}\left (x-\frac{y}{2} \right )  \psi_{N-1}\left (x+\frac{y}{2} \right )}{y}  \re^{ \frac{\ri p y}{\hbar} }.
 \ee
  In the first step, we replace $\psi_n(x)$ by its WKB approximation, which is given, in the classically allowed region, by 
 \be
\psi^{\rm WKB}_n (x) = \frac{1}{ \sqrt{2\pi p(x,E_n)} } \left ( \re^{\frac{\ri}{\hbar}S_n(x) +\frac{\ri \pi}{4} } + \re^{-\frac{\ri}{\hbar}S_n(x) -\frac{\ri \pi}{4} }  \right ), 
 \ee
 where we have denoted
  \be
 	p(x,E_n)=\sqrt{2 E_n-x^2},  \qquad S_n(x) = \int_{\sqrt{2 E_n}}^x p(x',E_n)\rd x'.
 \ee
 If we introduce the function
 \be
 \label{wmn}
 	w_{M,N}(x,p) = \int_{\mathbb R+\ri 0} \rd y \frac{ \psi^{\rm WKB}_M \left (x+\frac{y}{2} \right)\psi^{\rm WKB}_N \left (x-\frac{y}{2} \right)}{y}  \re^{ \frac{\ri p y}{\hbar} },
 \ee
we find that
\be
{\mathcal W}_{N}(x,p) \approx  \frac{1}{2\pi \sqrt{2 N \hbar}} \left( w_{N, N-1}(x,p)+{\text{h.c.}}\right). 
\ee
Furthermore, we are interested in the limit (\ref{ds-largeN}). To lighten the notation will write $\xi=N\hbar$. In this limit 
\be
\ba
	S_N(x) = {\mathcal S_0}(x)+\hbar {\mathcal S}_1(x)+\CO(\hbar^2), \\
	S_{N-1}(x) = {\mathcal S_0}(x)-\hbar {\mathcal S}_1(x)+\CO(\hbar^2), \\
\ea
\ee
where
\be
\ba
	{\mathcal S_0}(x) &= \frac{1}{2} x \sqrt{2\xi-x^2}+\ri \xi \log \left ( \frac{x+\ri \sqrt{2\xi-x^2} }{\sqrt{2\xi}} \right ), \\
	{\mathcal S_1}(x) &=\frac{\ri}{2} \log \left ( \frac{x+\ri \sqrt{2\xi-x^2} }{\sqrt{2\xi}} \right ),
\ea
\ee
and
\be
	p(x,E_{N}) = \sqrt{2\xi -x^2}+\CO(\hbar), \qquad p(x,E_{N-1})(x) = \sqrt{2\xi -x^2}+\CO(\hbar).
\ee
In the calculation of (\ref{wmn}) there are in principle four different terms. It can be seen that, in the saddle-point approximation, only one term contributes to the final result, 
and which one of the four terms contributes depends on the location of the point $(x,p)$. 
However, once the contribution from a single term is formulated geometrically, in terms of area of chords, the result is universal. Moreover, although 
we are doing the calculation in the classically allowed region, the result is valid everywhere, provided the area of the chord is analytically continued 
(see \cite{gm} for a detailed discussion of these issues in the context of Berry's original derivation of (\ref{berry-f})). In our case, it is enough to consider the term 
\be
\label{wn-int}
	w_{N,N-1}(x,p) \approx \frac{1}{2\pi} \int_{\mathbb R+\ri 0}\rd z  \frac{f(x-z) f(x+z) }{z} \re^{\frac{\ri}{\hbar}\Sigma(z)},
\ee
where we have rescaled $z\rightarrow 2z$. In this expression, we have 
\be
\ba
\Sigma(z) &= {\mathcal S_0}(x-z)-{\mathcal S_0}(x+z)+2p z, \\
f(x) &= \frac{1}{(2\xi)^{1/4}} \, \frac{\sqrt{x+\ri \sqrt{2\xi-x^2}}}{(2\xi-x^2)^{1/4}}.
\ea
\ee 
To perform the integral (\ref{wn-int}), we will use, as in \cite{berry}, the uniform saddle-point approximation of \cite{uniformsp}. We introduce 
a new integration variable $u$ (a uniformization variable) which satisfies
\be
\label{u-var}
	\Sigma(z) = \frac{u^3}{3} - \zeta u.
\ee
Since $\Sigma(z)$ is odd, the point $u=0$ satisfies $z=0$. The value of $\zeta$ can be obtained from the saddle points $\pm z^*$ satisfying 
\be
	0 = \Sigma'(\pm z_*) = -{\mathcal S_0}'(x + z_*) -{\mathcal S_0}'(x - z_*)+2p.
\ee
This is, with a slightly different notation, the condition (\ref{x0def}), which defines a chord passing through the point $(x, p)$ inside a circle of square radius $2 \xi$.  
After taking a derivative in (\ref{u-var}) and evaluating the result at the saddle points, one finds
\be
\label{zeta-chord}
\zeta = \left ( -\frac{3}{2} \Sigma(z_*) \right )^{2/3}  = \left ( \frac{3}{2} {\mathcal A(x,p)}\right )^{2/3},
\ee
where $\mathcal A(x,p)$ is the area of the Berry chord passing by the point $(x,p)$. It is given by:
\be
	\mathcal A(x,p) =2 \xi \left\{ \arccos \left( \sqrt{H (x,p) \over  \xi} \right) -\sqrt{H (x,p) \over  \xi}\sqrt{1-{H (x,p) \over  \xi}} \right\}, 
\ee
where $H(x,p)$ is the classical Hamiltonian (\ref{cho}). We now expand the remaining piece in the integrand of (\ref{wn-int}) as 
\be
	\label{expansionu}
	  \frac{f(x-z(u)) f(x+z(u)) }{z(u)} \frac{\rd z(u)}{\rd u} = \frac{a_{-1}}{u}+\sum_{m \geq 0} a_m u (u^2-\zeta)^m, 
\ee
where 
\be
	a_{-1} = f(x)^2 = \frac{1}{ \sqrt{2\xi}} \left ( \frac{x}{\sqrt{2\xi-x^2}}+\ri \right ).
\ee
If we keep just the first term in the expansion (\ref{expansionu}), we are left with the integral 
\be
	\frac{1}{2\pi}  \int_{\mathbb R+\ri 0} \frac{\rd u}{u} \, \re^{ \frac{\ri}{\hbar} (u^3/3-\zeta u) } = -\ri {\mathcal I}(-\hbar^{-2/3} \zeta), 
	\ee
	where $\CI$ is the integral of the Airy function introduced in (\ref{airy-int}). We conclude that 
\be
	w_{N,N-1}(x,p) \approx  -\ri  f(x)^2 \, \mathcal I \left (-\left ( \frac{3\mathcal A(x,p)}{2\hbar} \right )^{2/3}\right ), 
\ee
and 
\be
	\label{finalwigCNqho}
	{\mathcal W}_{N}(x,p) \approx \frac{1}{2\pi \xi} \, \mathcal I \left (-\left ( \frac{3\mathcal A(x,p)}{2\hbar} \right )^{2/3} \right ), 
\ee
which is the improved Balazs--Zipfel approximation. A comparison between the exact function ${\mathcal W}_{N}(x,p)$ and the improved Balazs--Zipfel approximation is shown in \figref{ibz-ho}. 
The agreement is very good. It is easy to verify that the improved Balazs--Zipfel approximation is closer to the exact result than the conventional Balazs--Zipfel approximation (in particular, 
it reproduces much better the pattern of fluctuations). 

It is easy to check that the neglected terms in (\ref{expansionu}) give an explicitly calculable series of corrections in powers of $\hbar^{2/3}$, involving Airy functions and their derivatives. After including the first correction, the approximation to $\mathcal W_{N}$ and the exact result can be hardly told apart, even for low values of $N$.

 \sectiono{Semiclassical distribution for local $\IF_0$}
 \label{app-localF0}
 In this Appendix we derive the approximation (\ref{bzFunc}) for the local $\IF_0$ geometry. The strategy is very similar to the one followed in the case of the harmonic oscillator. First, we use the results in section \ref{sec-red-mm} to write down the integral kernel of the reduced density matrix in the limit (\ref{sc-limit}). The resulting structure 
 is very similar to the integrand of a Wigner transform in the WKB approximation. Next, we evaluate the Wigner transform in the uniform saddle-point approximation.

Let us first calculate the integral kernel $C_N(x,y)$. In the case of local $\IF_0$, the constant 
$C$ in the integral kernel (\ref{rhoGen1}) vanishes, 
and one can use existing matrix model technology to evaluate (\ref{BNoZNanalytic}) explicitly. Since $\gamma=1$ for local $\IF_0$, the variable $M$ is given by $M=\re^\mu$. 
 As in similar calculations in \cite{kmz,mz-wv}, the $n$-point 
 correlation functions can be obtained from the spectral curve 
 \be
	\label{surfXY}
	(\re^\mu+\re^{-\mu})(\re^{\upsilon}+\re^{-\upsilon})-\kappa=0, 
\ee
where the modulus $\kappa$ is related to the 't Hooft parameter $\lambda$ through
\be
	\label{lambdaofkappa}
\lambda= \frac{\kappa}{16\pi^4} G_{3,3}^{2,3}\left(
\begin{array}{c}
 \frac{1}{2},\frac{1}{2},\frac{1}{2} \\
 0,0, -\frac{1}{2} \\
\end{array}
\left | \frac{\kappa^2}{16} \right.
\right) -{1\over 4 \pi}.
\ee
Therefore, by using (\ref{ed-lam}) and (\ref{ie-kappa}) we can identify $\kappa$ with the exponentiated, dual Fermi energy $\re^{E_{DF}}$. 
The elliptic modulus of the curve is 
\be
\label{el-mod}
	\tau =   \ri \frac{ {\rm K}(16/\kappa^2)}{ {\rm K}(1-16/\kappa^2)}. 
	\ee
The planar one-point function or resolvent $W_{1,0}(M)$ was already worked out in \cite{mz-wv}. If we solve for $\upsilon(\mu)$ as
\be
\upsilon(\mu) = \log \left ( \frac{\re^\mu \kappa+2 \ri \sqrt{\sigma(\re^\mu)} }{2(1+\re^{2\mu})} \right ), \
\ee
where
\be
	\sigma(M) = M^4-\frac{1}{4}M^2(\kappa^2-8)+1, 
\ee
then we have
\be
	W_{1,0}(M) = \frac{\ri}{2\pi^2 M} \upsilon (\mu)+\frac{1}{2 M}V_0'(\mu),
	\ee
where $V_0(\mu)$ is given in (\ref{vomu}). 
The planar two-point function was also written down in \cite{mz-wv}, by using the results of \cite{ek1}, and it is given by 
\be
\ba
	W_{2,0}(X_1,X_2) &= \frac{1}{2 \sqrt{\sigma(X_1)} \sqrt{\sigma(X_2)}}\left [a^2+b^2-2b^2\frac{ {\rm E}(1-\frac{a^2}{b^2}) }{ {\rm K}(1-\frac{a^2}{b^2}) }
	 \right . \\
	& \qquad  \qquad \qquad
	\left.
	 -(X_1^2+X_2^2) \left (1- \frac{ \left (\sqrt{\sigma(X_1)} - \sqrt{\sigma(X_2)} \right )^2 }{(X_1^2-X_2^2)^2}  \right ) \right ].
\ea
\ee
In this equation, $a$, $b$ are the endpoints of the cut where the eigenvalues condense, which are given in our case by 
\be
	b = 1/a = \frac{1}{4} \left (\kappa-\sqrt{\kappa^2-16} \right ).
\ee
The integrals of the planar two-point function can be written in terms of the Jacobi theta function $\vartheta_1(u)$ with modulus $\tau$ in (\ref{el-mod}), 
and the Abel--Jacobi map of the spectral curve, 
\be
	\label{AJexample}
	u(X) = c_\kappa \int_{\infty}^X \frac{1}{\sqrt{\sigma(X')}} \rd X', \qquad c_\kappa =   \frac{\kappa}{8 \ri {\rm K}(1-16/\kappa^2)}. 
	\ee
Indeed, one finds, for the integrals appearing in (\ref{BNoZNanalytic}), 
\be
	\int_{\infty}^{X_1} \int_{\infty}^{X_2} W_{2,0}(X_1',X_2')  \rd X_1' \rd X_2' = \log \left ( -\frac{ \vartheta_1(u(X_1)-u(X_2)) }{X_1-X_2} \frac{X_1+X_2}{ \vartheta_1(u(X_1)+u(X_2)) } \right ),
\ee
and
\be
	\int_{\infty}^{X} \int_{\infty}^{X} W_{2,0}(X_1',X_2')  \rd X_1' \rd X_2' = \frac{1}{2} \log \left [  \left ( -\frac{2 c_\kappa \vartheta_1'(0)}{\vartheta_1(2u(X))} \frac{X}{\sqrt{\sigma(X)}} \right )^2 \right ].
\ee
To write down the final result, let us define
\be
 {\cal D} = \re^{  \hbar \left ( \int_{\infty}^{\re^a} W_{1,0}(X) \rd X -\frac{1}{2}V_0(a) \right )}.
\ee
Then, one has
\be
	\label{Blargehbar}
\ba
	\frac{B_N(\mu_1,\mu_2)}{Z_N} &= \frac{2 c_\kappa \vartheta_1'(0){\cal D}^2}{2\pi} \re^{\frac{\ri \hbar}{2\pi^2} \int_a^{\mu_1} \upsilon (x)\rd x } \re^{\frac{\ri \hbar}{2\pi^2} \int_{a}^{\mu_2} \upsilon (x)\rd x } \frac{\re^{\mu_1} \re^{\mu_2}}{\re^{\mu_1}-\re^{\mu_2}}  \\
	& \quad \times \frac{1}{\sqrt{ \sqrt{\sigma(\re^{\mu_1})} \vartheta_1(2u(\re^{\mu_1}))} \sqrt{ \sqrt{\sigma(\re^{\mu_2})} \vartheta_1(2u(\re^{\mu_2}))}} \frac{ \vartheta_1(u(\re^{\mu_1})-u(\re^{\mu_2}))}{  \vartheta_1(u(\re^{\mu_1})+u(\re^{\mu_2})) } \left (1+\mathcal O(\hbar^{-1})\right ).
\ea
\ee
It turns out that this expression is valid as long as the variables $M_1$, $M_2$ do not belong to the interval where the cut occurs, namely $\CC=[a, b]$. When one of the variables is in the cut, 
it has to be modified as follows. The Riemann surface defined by (\ref{surfXY}) is a two-sheeted cover of the complex plane, and the two sheets correspond to the two sign determinations in front of the square-root when solving for $\upsilon(\mu)$.
 Let us define $\overline M_i =\exp(\bar \mu_i)$ as the point on the Riemann surface which corresponds to $M_i$, but on the other sheet. We have, in particular, 
 \be
	\label{analyticCont1}
\ba
	\sqrt{\sigma(\re^{\bar \mu})}&= -\sqrt{ \sigma(\re^{\mu})}, \\
	\upsilon(\bar \mu) &= -\upsilon (\mu) , \\
	u(\re^{\bar \mu}) &= 2u(\re^{a})-u(\re^{\mu}).
\ea
\ee
 We now define:
 \be
 	\label{BoZana2}
 \ba
 	B(\mu_1,\mu_2) = 
	\begin{cases}
		\frac{B_N(\mu_1,\mu_2)}{Z_N} & \qquad \text{for } M_1,M_2 \notin {\mathcal C} \\
		\frac{B_N(\mu_1,\mu_2)}{Z_N} +\frac{B_N(\bar \mu_1,\mu_2)}{Z_N}& \qquad \text{for } M_1 \in {\mathcal C},M_2 \notin {\mathcal C} \\
		\frac{B_N(\mu_1,\mu_2)}{Z_N} +\frac{B_N(\mu_1,\bar \mu_2)}{Z_N}& \qquad \text{for } M_1 \notin {\mathcal C},M_2 \in {\mathcal C} \\
		\frac{B_N(\mu_1,\mu_2)}{Z_N} +\frac{B_N(\bar \mu_1,\mu_2)}{Z_N}+\frac{B_N( \mu_1, \bar \mu_2)}{Z_N} + \frac{B_N(\bar \mu_1, \bar \mu_2)}{Z_N} & \qquad \text{for } M_1,M_2 \in {\mathcal C}. \\
	\end{cases}
 \ea
 \ee
We have explicitly verified that this function gives an excellent approximation to the reduced density matrix (which can be also evaluated numerically).\footnote{The fact that we need to evaluate the function $B_N(\mu_1,\mu_2)$ on its different sheets when $\re^{\mu_i}$ is inside the cut $\mathcal C$ can be understood from the large $N$ matrix model. Indeed, when $\re^{\mu_i}$ is in $\mathcal C$, this means that $\mu_i$ is inside the interval where the eigenvalues of the matrix model condense in the large $N$ limit. In that case, large $N$ expectation values in the matrix model (such as $B_N(\mu_1,\mu_2)$) are ambiguous due to the presence of the branch cut, and taking the average above and below the cut is the usual prescription.}

We can now use the analytic expression (\ref{Blargehbar}) to calculate $\CQ_N(x,p)$ in the semiclassical limit (\ref{sc-limit}). One finds, 
\be
\CQ_{N+1} (x, p) \approx \frac{Z_N}{(N+1)Z_{N+1}} \frac{1}{2\pi \hbar_D} \int_{\mathbb R} \rd z  \, \re^{ \ri (x+p) z/\hbar_D} \, B \left (\frac{x- p}{2}-\frac{z}{2}, \frac{ x- p}{2}+\frac{z}{2} \right).
\ee
In terms of the previous variables, we have
\be
	\mu= \frac{x-p}{2}, \qquad  \upsilon = \frac{x+p}{2}, 
\ee
and $\upsilon$ is related to $\mu$ by the equation of the curve (\ref{surfXY}). As in the case of the harmonic oscillator discussed in Appendix \ref{ho}, we consider the analogue of the 
classically allowed region, which is the case with $M_{1,2} \in \CC$. This is, as shown in (\ref{BoZana2}), the sum of four terms. As in the case of the harmonic oscillator, it is enough to use the term which leads 
straightforwardly to the chord construction, which is $B(\mu_1, \bar \mu_2)$. One obtains the following expression (after rescaling $z\rightarrow 2z$):
\be
	\label{intQClargehbar}
	\CQ_{N+1} (x,p) \approx -\ri \mathcal N \int_{\mathbb R+\ri0} \rd z\, \re^{2 \ri\Sigma(z)/\hbar_D }\frac{F(z)}{\re^{\mu-z}-\re^{\mu+z}},
\ee
where
\be
\ba
	\Sigma(z) &= 2 \upsilon z + \int_a^{\mu-z} \upsilon (\mu')\rd \mu' - \int_a^{\mu+z}\upsilon (\mu')\rd \mu',
	\\
	F(z) &= \frac{ \re^{2\mu} }{ \sqrt{\sqrt {\sigma(\re^{\mu-z})} \vartheta_1(2 u(\re^{\mu-z})) }  \sqrt{-\sqrt {\sigma(\re^{\mu+z})} \vartheta_1(4 u(\re^{a})-2 u(\re^{\mu+z})) }}  \\
	& \qquad \qquad \qquad \qquad \qquad \qquad
	\times 
	\frac{\vartheta_1(u(\re^{\mu-z})+u(\re^{\mu+z})-2 u(\re^a))}{ \vartheta_1(u(\re^{\mu-z})-u(\re^{\mu+z})+2 u(\re^a)) },
	\\
	\mathcal N &=  \frac{ 2 Z_N}{(N+1)Z_{N+1}} \cdot  \frac{1}{2\pi \hbar_D} \cdot \frac{2\ri c_\kappa {\vartheta_1'(0)}{\cal D}^2}{2\pi}.
\ea
\ee
We can now perform the integral in (\ref{intQClargehbar}) by using the uniform saddle-point approximation, as in (\ref{wn-int}). In view of the form of $\Sigma(z)$, this will involve again Berry's chord 
construction. If we introduce a uniformization variable $u$ as in (\ref{u-var}), the value of $\zeta$ is given by 
\be
\zeta= \left( { 3  \over 4} \CA(x,p)  \right)^{2/3}, 
\ee
where $\CA(x,p) $ is the area of the chord associated to the curve 
(\ref{dmc}) with $\CO(\re^x, \re^p)$ given in (\ref{mc-f0}) (the extra factor of $1/2$ w.r.t. (\ref{zeta-chord}) is due to the fact that we have parametrized $\Sigma(z)$ with the variables $\mu$, $\upsilon$, and 
the volume form in these variables is $1/2$ of the volume form in the variables $x, p$). We now expand the integrand of (\ref{intQClargehbar}) as in (\ref{expansionu}), 
\be
	\label{insideIntexp1}
	  \frac{F(z(u))}{\re^{\mu-z(u)}-\re^{\mu+z(u)}} z'(u) = \frac{p_{-1}}{u}+\sum_{m \geq 0} p_m u (u^2-\zeta)^m+\sum_{m \geq 0} q_m (u^2-\zeta)^m.
\ee
The leading term comes from $p_{-1}$, which is given by 
\be
	p_{-1}=- \frac{1}{2} F(0)\re^{-\mu} 
\ee
and turns out to be a constant, equal to 
\be
  -\frac{\ri \re^{\ri \pi \tau/2} }{4 c_{\kappa} \vartheta_1'(0)}. 
  \ee
We finally obtain, 
\be
\ba
	\CQ_{N+1} (x,p) &\approx -\mathcal N \pi \left (  \frac{\ri \re^{\ri \pi \tau} }{2 c_{\kappa} \vartheta_1'(0)} \right ) {\mathcal I} \left ( - \left ( \frac{2}{\hbar_D}\right)^{2/3} \zeta  \right )\\
&\approx  \frac{1}{2\pi \hbar_D(N+1)} {\mathcal I} \left ( - \left ( \frac{3}{2 \hbar_D} \mathcal A (x,p) \right)^{2/3}   \right ).
\ea
\ee
In the last step, we used the 't Hooft expansion of $Z_N$ and $Z_{N+1}$, as well as some identities coming from the special geometry of local $\IF_0$. The final result is the improved Balazs--Zipfel 
approximation (\ref{bzFunc}), with a small caveat: in the r.h.s., the 't Hooft parameter governing the shape of the mirror curve is determined by $N$, and not by $N+1$. This however gives a subleading correction 
to the asymptotics for large $N$. We have also verified that the next-to-leading correction to our formula has approximately the effect of shifting $N$ by one unit.

\bibliographystyle{JHEP}

\linespread{0.6}
\bibliography{biblio}

\providecommand{\href}[2]{#2}\begingroup\raggedright\begin{thebibliography}{10}

\bibitem{hillery}
M.~Hillery, R.~F. O'Connell, M.~O. Scully and E.~P. Wigner, \emph{{Distribution
  functions in physics: Fundamentals}},
  \href{http://dx.doi.org/10.1016/0370-1573(84)90160-1}{\emph{Phys. Rept.} {\bf
  106} (1984) 121--167}.

\bibitem{zachos-treatise}
T.~L. Curtright, D.~B. Fairlie and C.~K. Zachos, \emph{A concise treatise on
  quantum mechanics in phase space}.
\newblock World Scientific Publishing Company, 2013.

\bibitem{halliwell}
J.~J. Halliwell, \emph{{Correlations in the Wave Function of the Universe}},
  \href{http://dx.doi.org/10.1103/PhysRevD.36.3626}{\emph{Phys. Rev.} {\bf D36}
  (1987) 3626}.

\bibitem{anderson}
A.~Anderson, \emph{{On Predicting Correlations From Wigner Functions}},
  \href{http://dx.doi.org/10.1103/PhysRevD.42.585}{\emph{Phys. Rev.} {\bf D42}
  (1990) 585--589}.

\bibitem{habib}
S.~Habib, \emph{{The classical limit in quantum cosmology. 1 Quantum mechanics
  and the Wigner function}},
  \href{http://dx.doi.org/10.1103/PhysRevD.42.2566}{\emph{Phys. Rev.} {\bf D42}
  (1990) 2566--2576}.

\bibitem{hh}
J.~B. Hartle and S.~W. Hawking, \emph{{Wave Function of the Universe}},
  \href{http://dx.doi.org/10.1103/PhysRevD.28.2960}{\emph{Phys. Rev.} {\bf D28}
  (1983) 2960--2975}.

\bibitem{gomez}
C.~G\'omez, S.~Mont\'a\~nez and P.~Resco, \emph{{Semi-classical mechanics in
  phase space: The Quantum target of minimal strings}},
  \href{http://dx.doi.org/10.1088/1126-6708/2005/11/049}{\emph{JHEP} {\bf 11}
  (2005) 049}, [\href{http://arxiv.org/abs/hep-th/0506159}{{\tt
  hep-th/0506159}}].

\bibitem{ambjorn}
J.~Ambjorn and R.~A. Janik, \emph{{The emergence of noncommutative target space
  in noncritical string theory}},
  \href{http://dx.doi.org/10.1088/1126-6708/2005/08/057}{\emph{JHEP} {\bf 08}
  (2005) 057}, [\href{http://arxiv.org/abs/hep-th/0506197}{{\tt
  hep-th/0506197}}].

\bibitem{moore-rs}
G.~W. Moore, \emph{{Geometry of the string equations}},
  \href{http://dx.doi.org/10.1007/BF02097368}{\emph{Commun. Math. Phys.} {\bf
  133} (1990) 261--304}.

\bibitem{mmss}
J.~M. Maldacena, G.~W. Moore, N.~Seiberg and D.~Shih, \emph{{Exact vs.
  semiclassical target space of the minimal string}},
  \href{http://dx.doi.org/10.1088/1126-6708/2004/10/020}{\emph{JHEP} {\bf 0410}
  (2004) 020}, [\href{http://arxiv.org/abs/hep-th/0408039}{{\tt
  hep-th/0408039}}].

\bibitem{dmw1}
A.~Dhar, G.~Mandal and S.~R. Wadia, \emph{{Classical Fermi fluid and geometric
  action for c=1}},
  \href{http://dx.doi.org/10.1142/S0217751X93000138}{\emph{Int. J. Mod. Phys.}
  {\bf A8} (1993) 325--350}, [\href{http://arxiv.org/abs/hep-th/9204028}{{\tt
  hep-th/9204028}}].

\bibitem{dmw2}
A.~Dhar, G.~Mandal and S.~R. Wadia, \emph{{Nonrelativistic fermions, coadjoint
  orbits of W(infinity) and string field theory at c = 1}},
  \href{http://dx.doi.org/10.1142/S0217732392002512}{\emph{Mod. Phys. Lett.}
  {\bf A7} (1992) 3129--3146}, [\href{http://arxiv.org/abs/hep-th/9207011}{{\tt
  hep-th/9207011}}].

\bibitem{berenstein}
D.~Berenstein, \emph{{A Toy model for the AdS / CFT correspondence}},
  \href{http://dx.doi.org/10.1088/1126-6708/2004/07/018}{\emph{JHEP} {\bf 07}
  (2004) 018}, [\href{http://arxiv.org/abs/hep-th/0403110}{{\tt
  hep-th/0403110}}].

\bibitem{llm}
H.~Lin, O.~Lunin and J.~M. Maldacena, \emph{{Bubbling AdS space and 1/2 BPS
  geometries}},
  \href{http://dx.doi.org/10.1088/1126-6708/2004/10/025}{\emph{JHEP} {\bf 10}
  (2004) 025}, [\href{http://arxiv.org/abs/hep-th/0409174}{{\tt
  hep-th/0409174}}].

\bibitem{babel}
V.~Balasubramanian, J.~de~Boer, V.~Jejjala and J.~Sim\'on, \emph{{The Library
  of Babel: On the origin of gravitational thermodynamics}},
  \href{http://dx.doi.org/10.1088/1126-6708/2005/12/006}{\emph{JHEP} {\bf 12}
  (2005) 006}, [\href{http://arxiv.org/abs/hep-th/0508023}{{\tt
  hep-th/0508023}}].

\bibitem{st}
K.~Skenderis and M.~Taylor, \emph{{Anatomy of bubbling solutions}},
  \href{http://dx.doi.org/10.1088/1126-6708/2007/09/019}{\emph{JHEP} {\bf 09}
  (2007) 019}, [\href{http://arxiv.org/abs/0706.0216}{{\tt 0706.0216}}].

\bibitem{joan1}
V.~Balasubramanian, B.~Czech, K.~Larjo and J.~Sim\'on, \emph{{Integrability
  versus information loss: A Simple example}},
  \href{http://dx.doi.org/10.1088/1126-6708/2006/11/001}{\emph{JHEP} {\bf 11}
  (2006) 001}, [\href{http://arxiv.org/abs/hep-th/0602263}{{\tt
  hep-th/0602263}}].

\bibitem{joan2}
V.~Balasubramanian, B.~Czech, K.~Larjo, D.~Marolf and J.~Sim\'on,
  \emph{{Quantum geometry and gravitational entropy}},
  \href{http://dx.doi.org/10.1088/1126-6708/2007/12/067}{\emph{JHEP} {\bf 12}
  (2007) 067}, [\href{http://arxiv.org/abs/0705.4431}{{\tt 0705.4431}}].

\bibitem{adkmv}
M.~Aganagic, R.~Dijkgraaf, A.~Klemm, M.~Mari\~no and C.~Vafa,
  \emph{{Topological strings and integrable hierarchies}},
  \href{http://dx.doi.org/10.1007/s00220-005-1448-9}{\emph{Commun.Math.Phys.}
  {\bf 261} (2006) 451--516}, [\href{http://arxiv.org/abs/hep-th/0312085}{{\tt
  hep-th/0312085}}].

\bibitem{ghm}
A.~Grassi, Y.~Hatsuda and M.~Mari\~no, \emph{{Topological strings from quantum
  mechanics}}, \href{http://dx.doi.org/10.1007/s00023-016-0479-4}{\emph{Annales
  Henri Poincar\'e} {\bf 17} (2016) 3177--3235},
  [\href{http://arxiv.org/abs/1410.3382}{{\tt 1410.3382}}].

\bibitem{mz}
M.~Mari\~no and S.~Zakany, \emph{{Matrix models from operators and topological
  strings}}, \href{http://dx.doi.org/10.1007/s00023-015-0422-0}{\emph{Annales
  Henri Poincar\'e} {\bf 17} (2016) 1075--1108},
  [\href{http://arxiv.org/abs/1502.02958}{{\tt 1502.02958}}].

\bibitem{dm1}
O.~Dumitrescu and M.~Mulase, \emph{{Lectures on the topological recursion for
  Higgs bundles and quantum curves}},
  \href{http://arxiv.org/abs/1509.09007}{{\tt 1509.09007}}.

\bibitem{dm2}
O.~Dumitrescu and M.~Mulase, \emph{{An invitation to 2D TQFT and quantization
  of Hitchin spectral curves}},  \href{http://arxiv.org/abs/1705.05969}{{\tt
  1705.05969}}.

\bibitem{be-wkb}
V.~Bouchard and B.~Eynard, \emph{{Reconstructing WKB from topological
  recursion}},  \href{http://arxiv.org/abs/1606.04498}{{\tt 1606.04498}}.

\bibitem{ms}
M.~Manabe and P.~Su{\l}kowski, \emph{{Quantum curves and conformal field
  theory}}, \href{http://dx.doi.org/10.1103/PhysRevD.95.126003}{\emph{Phys.
  Rev.} {\bf D95} (2017) 126003}, [\href{http://arxiv.org/abs/1512.05785}{{\tt
  1512.05785}}].

\bibitem{wzh}
X.~Wang, G.~Zhang and M.-x. Huang, \emph{{New exact quantization condition for
  toric Calabi--Yau geometries}},
  \href{http://dx.doi.org/10.1103/PhysRevLett.115.121601}{\emph{Phys. Rev.
  Lett.} {\bf 115} (2015) 121601}, [\href{http://arxiv.org/abs/1505.05360}{{\tt
  1505.05360}}].

\bibitem{hatsuda-comments}
Y.~Hatsuda, \emph{{Comments on exact quantization conditions and
  non-perturbative topological strings}},
  \href{http://arxiv.org/abs/1507.04799}{{\tt 1507.04799}}.

\bibitem{faddeev}
L.~Faddeev, \emph{{Discrete Heisenberg-Weyl group and modular group}},
  \href{http://dx.doi.org/10.1007/BF01872779}{\emph{Lett.Math.Phys.} {\bf 34}
  (1995) 249--254}, [\href{http://arxiv.org/abs/hep-th/9504111}{{\tt
  hep-th/9504111}}].

\bibitem{berry}
M.~V. Berry, \emph{{Semi-Classical Mechanics in Phase Space: A Study of
  Wigner's Function}},
  \href{http://dx.doi.org/10.1098/rsta.1977.0145}{\emph{Phil. Trans. Roy. Soc.
  Lond.} {\bf A287} (1977) 237--271}.

\bibitem{bzipfel}
N.~Balazs and G.~Zipfel~Jr, \emph{Quantum oscillations in the semiclassical
  fermion $\mu$-space density}, {\emph{Annals of Physics} {\bf 77} (1973)
  139--156}.

\bibitem{negele-orland}
J.~W. Negele and H.~Orland, \emph{Quantum many-particle systems}.
\newblock Westview, 1988.

\bibitem{dlms}
D.~S. Dean, P.~Le~Doussal, S.~N. Majumdar and G.~Schehr, \emph{{Noninteracting
  fermions at finite temperature in a $d$-dimensional trap: Universal
  correlations}},
  \href{http://dx.doi.org/10.1103/PhysRevA.94.063622}{\emph{Phys. Rev.} {\bf
  A94} (2016) 063622}, [\href{http://arxiv.org/abs/1609.04366}{{\tt
  1609.04366}}].

\bibitem{krauth}
W.~Krauth, \emph{Statistical mechanics: algorithms and computations}.
\newblock Oxford University Press, Oxford, 2006.

\bibitem{mz-wv}
M.~Mari\~no and S.~Zakany, \emph{{Exact eigenfunctions and the open topological
  string}}, \href{http://dx.doi.org/10.1088/1751-8121/aa791e}{\emph{J. Phys.}
  {\bf A50} (2017) 325401}, [\href{http://arxiv.org/abs/1606.05297}{{\tt
  1606.05297}}].

\bibitem{voros-asymptotic}
A.~Voros, \emph{Asymptotic $\hbar$-expansions of stationary quantum states},
  {\emph{Ann. Inst. H. Poincar{\'e} A} {\bf 26} (1977) 343--403}.

\bibitem{ripamonti}
N.~Ripamonti, \emph{Classical limit of the harmonic oscillator {W}igner
  functions in the {B}argmann representation}, {\emph{Journal of Physics A:
  Mathematical and General} {\bf 29} (1996) 5137}.

\bibitem{dlms2}
D.~S. Dean, P.~Le~Doussal, S.~N. Majumdar and G.~Schehr, \emph{{Wigner function
  of noninteracting trapped fermions}},
  \href{http://arxiv.org/abs/1801.02680}{{\tt 1801.02680}}.

\bibitem{ns}
N.~A. Nekrasov and S.~L. Shatashvili, \emph{{Quantization of integrable systems
  and four dimensional gauge theories}},  in \emph{{16th International Congress
  on Mathematical Physics, Prague, August 2009, 265-289, World Scientific
  2010}}, 2009.
\newblock \href{http://arxiv.org/abs/0908.4052}{{\tt 0908.4052}}.

\bibitem{km}
J.~Kallen and M.~Mari\~no, \emph{{Instanton effects and quantum spectral
  curves}}, \href{http://dx.doi.org/10.1007/s00023-015-0421-1}{\emph{Annales
  Henri Poincar\'e} {\bf 17} (2016) 1037--1074},
  [\href{http://arxiv.org/abs/1308.6485}{{\tt 1308.6485}}].

\bibitem{mmrev}
M.~Mari\~no, \emph{{Spectral theory and mirror symmetry}},
  \href{http://arxiv.org/abs/1506.07757}{{\tt 1506.07757}}.

\bibitem{cgum}
S.~Codesido, J.~Gu and M.~Mari\~no, \emph{{Operators and higher genus mirror
  curves}}, \href{http://dx.doi.org/10.1007/JHEP02(2017)092}{\emph{JHEP} {\bf
  02} (2017) 092}, [\href{http://arxiv.org/abs/1609.00708}{{\tt 1609.00708}}].

\bibitem{ggu}
A.~Grassi and J.~Gu, \emph{{BPS relations from spectral problems and blowup
  equations}},  \href{http://arxiv.org/abs/1609.05914}{{\tt 1609.05914}}.

\bibitem{swh}
K.~Sun, X.~Wang and M.-x. Huang, \emph{{Exact Quantization Conditions, Toric
  Calabi-Yau and Nonperturbative Topological String}},
  \href{http://dx.doi.org/10.1007/JHEP01(2017)061}{\emph{JHEP} {\bf 01} (2017)
  061}, [\href{http://arxiv.org/abs/1606.07330}{{\tt 1606.07330}}].

\bibitem{h-blowup}
M.-x. Huang, K.~Sun and X.~Wang, \emph{{Blowup Equations for Refined
  Topological Strings}},  \href{http://arxiv.org/abs/1711.09884}{{\tt
  1711.09884}}.

\bibitem{acdkv}
M.~Aganagic, M.~C. Cheng, R.~Dijkgraaf, D.~Krefl and C.~Vafa, \emph{{Quantum
  geometry of refined topological strings}},
  \href{http://dx.doi.org/10.1007/JHEP11(2012)019}{\emph{JHEP} {\bf 1211}
  (2012) 019}, [\href{http://arxiv.org/abs/1105.0630}{{\tt 1105.0630}}].

\bibitem{hw}
M.-x. Huang and X.-f. Wang, \emph{{Topological strings and quantum spectral
  problems}}, \href{http://dx.doi.org/10.1007/JHEP09(2014)150}{\emph{JHEP} {\bf
  1409} (2014) 150}, [\href{http://arxiv.org/abs/1406.6178}{{\tt 1406.6178}}].

\bibitem{bipz}
E.~Br\'ezin, C.~Itzykson, G.~Parisi and J.~B. Zuber, \emph{{Planar Diagrams}},
  \href{http://dx.doi.org/10.1007/BF01614153}{\emph{Commun. Math. Phys.} {\bf
  59} (1978) 35}.

\bibitem{kmz}
R.~Kashaev, M.~Mari\~no and S.~Zakany, \emph{{Matrix models from operators and
  topological strings, 2}},
  \href{http://dx.doi.org/10.1007/s00023-016-0471-z}{\emph{Annales Henri
  Poincar\'e} {\bf 17} (2016) 2741--2781},
  [\href{http://arxiv.org/abs/1505.02243}{{\tt 1505.02243}}].

\bibitem{kaserg}
R.~M. Kashaev and S.~M. Sergeev, \emph{{Spectral equations for the modular
  oscillator}},  \href{http://arxiv.org/abs/1703.06016}{{\tt 1703.06016}}.

\bibitem{butterfly}
Y.~Hatsuda, H.~Katsura and Y.~Tachikawa, \emph{{Hofstadter's butterfly in
  quantum geometry}},
  \href{http://dx.doi.org/10.1088/1367-2630/18/10/103023}{\emph{New J. Phys.}
  {\bf 18} (2016) 103023}, [\href{http://arxiv.org/abs/1606.01894}{{\tt
  1606.01894}}].

\bibitem{gm-complex}
A.~Grassi and M.~Mari\~no, \emph{{The complex side of the TS/ST
  correspondence}},  \href{http://arxiv.org/abs/1708.08642}{{\tt 1708.08642}}.

\bibitem{Sciarappa-2}
A.~Sciarappa, \emph{{Exact relativistic Toda chain eigenfunctions from
  Separation of Variables and gauge theory}},
  \href{http://dx.doi.org/10.1007/JHEP10(2017)116}{\emph{JHEP} {\bf 10} (2017)
  116}, [\href{http://arxiv.org/abs/1706.05142}{{\tt 1706.05142}}].

\bibitem{gu-s}
J.~Gu and T.~Sulejmanpasic, \emph{{High order perturbation theory for
  difference equations and Borel summability of quantum mirror curves}},
  \href{http://dx.doi.org/10.1007/JHEP12(2017)014}{\emph{JHEP} {\bf 12} (2017)
  014}, [\href{http://arxiv.org/abs/1709.00854}{{\tt 1709.00854}}].

\bibitem{cms-np}
S.~Codesido, M.~Mari\~no and R.~Schiappa, \emph{{Non-Perturbative Quantum
  Mechanics from Non-Perturbative Strings}},
  \href{http://arxiv.org/abs/1712.02603}{{\tt 1712.02603}}.

\bibitem{entropy}
P.~Calabrese, P.~Le~Doussal and S.~N. Majumdar, \emph{Random matrices and
  entanglement entropy of trapped fermi gases}, {\emph{Physical Review A} {\bf
  91} (2015) 012303}.

\bibitem{kama}
R.~Kashaev and M.~Mari\~no, \emph{{Operators from mirror curves and the quantum
  dilogarithm}},
  \href{http://dx.doi.org/10.1007/s00220-015-2499-1}{\emph{Commun. Math. Phys.}
  {\bf 346} (2016) 967}, [\href{http://arxiv.org/abs/1501.01014}{{\tt
  1501.01014}}].

\bibitem{kwy-ids}
A.~Kapustin, B.~Willett and I.~Yaakov, \emph{{Nonperturbative Tests of
  Three-Dimensional Dualities}},
  \href{http://dx.doi.org/10.1007/JHEP10(2010)013}{\emph{JHEP} {\bf 1010}
  (2010) 013}, [\href{http://arxiv.org/abs/1003.5694}{{\tt 1003.5694}}].

\bibitem{mp}
M.~Mari\~no and P.~Putrov, \emph{{ABJM theory as a Fermi gas}},
  \href{http://dx.doi.org/10.1088/1742-5468/2012/03/P03001}{\emph{J.Stat.Mech.}
  {\bf 1203} (2012) P03001}, [\href{http://arxiv.org/abs/1110.4066}{{\tt
  1110.4066}}].

\bibitem{kopss}
D.~Kutasov, K.~Okuyama, J.-w. Park, N.~Seiberg and D.~Shih, \emph{{Annulus
  amplitudes and ZZ branes in minimal string theory}},
  \href{http://dx.doi.org/10.1088/1126-6708/2004/08/026}{\emph{JHEP} {\bf 08}
  (2004) 026}, [\href{http://arxiv.org/abs/hep-th/0406030}{{\tt
  hep-th/0406030}}].

\bibitem{ackm}
J.~Ambj{\o}rn, L.~Chekhov, C.~F. Kristjansen and {\relax Yu}.~Makeenko,
  \emph{{Matrix model calculations beyond the spherical limit}},
  \href{http://dx.doi.org/10.1016/0550-3213(93)90476-6,
  10.1016/0550-3213(95)00391-5}{\emph{Nucl. Phys.} {\bf B404} (1993) 127--172},
  [\href{http://arxiv.org/abs/hep-th/9302014}{{\tt hep-th/9302014}}].

\bibitem{ek1}
B.~Eynard and C.~Kristjansen, \emph{{Exact solution of the O(n) model on a
  random lattice}},
  \href{http://dx.doi.org/10.1016/0550-3213(95)00469-9}{\emph{Nucl. Phys.} {\bf
  B455} (1995) 577--618}, [\href{http://arxiv.org/abs/hep-th/9506193}{{\tt
  hep-th/9506193}}].

\bibitem{ek2}
B.~Eynard and C.~Kristjansen, \emph{{More on the exact solution of the $O(n)$
  model on a random lattice and an investigation of the case $|n|> 2$}},
  \href{http://dx.doi.org/10.1016/0550-3213(96)00104-6}{\emph{Nucl. Phys.} {\bf
  B466} (1996) 463--487}, [\href{http://arxiv.org/abs/hep-th/9512052}{{\tt
  hep-th/9512052}}].

\bibitem{garou-kas}
S.~Garoufalidis and R.~Kashaev, \emph{{Evaluation of state integrals at
  rational points}},
  \href{http://dx.doi.org/10.4310/CNTP.2015.v9.n3.a3}{\emph{Commun. Num. Theor.
  Phys.} {\bf 09} (2015) 549--582}, [\href{http://arxiv.org/abs/1411.6062}{{\tt
  1411.6062}}].

\bibitem{cfz}
T.~Curtright, D.~Fairlie and C.~K. Zachos, \emph{Features of time independent
  {W}igner functions},
  \href{http://dx.doi.org/10.1103/PhysRevD.58.025002}{\emph{Phys. Rev.} {\bf
  D58} (1998) 025002}, [\href{http://arxiv.org/abs/hep-th/9711183}{{\tt
  hep-th/9711183}}].

\bibitem{b-moyal}
M.~Bartlett and J.~Moyal, \emph{The exact transition probabilities of
  quantum-mechanical oscillators calculated by the phase-space method},
  {\emph{Mathematical Proceedings of the Cambridge Philosophical Society} {\bf
  45} (1949) 545--553}.

\bibitem{cgm}
S.~Codesido, A.~Grassi and M.~Mari\~no, \emph{{Spectral theory and mirror
  curves of higher genus}},
  \href{http://dx.doi.org/10.1007/s00023-016-0525-2}{\emph{Annales Henri
  Poincar\'e} {\bf 18} (2017) 559--622},
  [\href{http://arxiv.org/abs/1507.02096}{{\tt 1507.02096}}].

\bibitem{voros-husimi}
A.~Voros, \emph{{The {WKB} Method in the Bargmann Representation}},
  \href{http://dx.doi.org/10.1103/PhysRevA.40.6814}{\emph{Phys. Rev.} {\bf A40}
  (1989) 6814--6825}.

\bibitem{saraceno}
J.~Kurchan, P.~Leboeuf and M.~Saraceno, \emph{Semiclassical approximations in
  the coherent-state representation}, {\emph{Physical Review A} {\bf 40} (1989)
  6800}.

\bibitem{hartnoll}
S.~A. Hartnoll and E.~Mazenc, \emph{{Entanglement entropy in two dimensional
  string theory}},
  \href{http://dx.doi.org/10.1103/PhysRevLett.115.121602}{\emph{Phys. Rev.
  Lett.} {\bf 115} (2015) 121602}, [\href{http://arxiv.org/abs/1504.07985}{{\tt
  1504.07985}}].

\bibitem{gm}
K.-S. Giannopoulou and G.~N. Makrakis, \emph{An approximate series solution of
  the semiclassical {W}igner equation},
  \href{http://arxiv.org/abs/1705.06754}{{\tt 1705.06754}}.

\bibitem{uniformsp}
C.~Chester, B.~Friedman and F.~Ursell, \emph{An extension of the method of
  steepest descents},  in \emph{Mathematical Proceedings of the Cambridge
  Philosophical Society}, vol.~53, pp.~599--611, Cambridge University Press,
  1957.

\end{thebibliography}\endgroup
\end{document}